\documentclass[twocolumn,showpacs,preprintnumbers,aps,psfig,superscriptaddress]{revtex4-1}
\usepackage{amsmath}
\usepackage{graphicx}
\usepackage{dcolumn}
\usepackage{float}

\begin{document}
\title{Investigations of the singlet ground state system: PrIrSi$_{3}$}
\author{V. K. Anand}
\altaffiliation{vivekkranand@gmail.com}
\affiliation {ISIS Facility, Rutherford Appleton Laboratory, Chilton, Didcot, Oxon, OX11 0QX, United Kingdom}
\affiliation{Helmholtz-Zentrum Berlin f\"{u}r Materialien und Energie, Hahn-Meitner Platz 1, 14109 Berlin, Germany.}
\author{D. T. Adroja}
\altaffiliation{devashibhai.adroja@stfc.ac.uk}
\affiliation {ISIS Facility, Rutherford Appleton Laboratory, Chilton, Didcot, Oxon, OX11 0QX,  United Kingdom}
\affiliation {Physics Department, Highly Correlated Matter Research Group, University of Johannesburg, P.O. Box 524, Auckland Park 2006, South Africa}
\author{A. Bhattacharyya}
\affiliation {ISIS Facility, Rutherford Appleton Laboratory, Chilton, Didcot, Oxon, OX11 0QX,  United Kingdom}
\affiliation {Physics Department, Highly Correlated Matter Research Group, University of Johannesburg, P.O. Box 524, Auckland Park 2006, South Africa}
\author{A. D. Hillier}
\affiliation {ISIS Facility, Rutherford Appleton Laboratory, Chilton, Didcot, Oxon, OX11 0QX,  United Kingdom}
\author{J. W. Taylor}
\affiliation {ISIS Facility, Rutherford Appleton Laboratory, Chilton, Didcot, Oxon, OX11 0QX,  United Kingdom}
\author{A. M. Strydom}
\affiliation {Physics Department, Highly Correlated Matter Research Group, University of Johannesburg, P.O. Box 524, Auckland Park 2006, South Africa}

\date{\today}

\begin{abstract}
We report our comprehensive study of physical properties of a ternary intermetallic compound PrIrSi$_{3}$ investigated by dc magnetic susceptibility $\chi(T)$, isothermal magnetization $M(H)$, thermo-remnant magnetization $M(t)$, ac magnetic susceptibility $\chi_{\rm ac}(T)$, specific heat $C_{\rm p}(T)$, electrical resistivity $\rho(T)$, muon spin relaxation ($\mu$SR) and inelastic neutron scattering (INS) measurements. A magnetic phase transition is marked by a sharp anomaly at $T_{tr} = 12.2$~K in $\chi(T)$ measured at low applied fields which is also reflected in the $C_{\rm p}(T)$ data through a weak anomaly at 12~K\@. An irreversibility between the zero field cooled and field cooled $\chi(T)$ data below 12.2~K and a very large relaxation time of $M(t)$ indicates the presence of ferromagnetic correlation. The magnetic part of specific heat shows a broad Schottky-type anomaly near 40~K due to the crystal electric field (CEF) effect. An extremely small value of magnetic entropy below 12~K suggests a CEF-split singlet ground state which is confirmed from our analysis of INS data. The INS spectra show two prominent inelastic excitations at 8.5~meV and 18.5~meV that could be well accounted by a CEF model. The CEF splitting energy between the ground state singlet and the first excited doublet is found to be 92~K\@. Our $\mu$SR data reveal a possible magnetic ordering below 30~K which is much higher than that found from the specific heat and magnetic susceptibility measurements. This could be due to the presence of short range correlations well above the long range magnetic ordering or due to the electronic changes induced by muons. The induced moment magnetism in the singlet ground state system PrIrSi$_{3}$ with such a large splitting energy of 92~K is quite surprising.
\end{abstract}

\pacs{71.70.-d, 78.70.Nx, 76.75.+i, 75.30.-m}

\maketitle

\section{\label{Intro} INTRODUCTION}

The electrostatic coupling between the $4f$ shell of rare earths having nonzero orbital angular momentum ($L \neq 0$) and its environment, known as crystalline electric field (CEF), greatly influences the physical properties of a rare earth system. The CEF modifies the energy levels of rare earth atoms in a solid and tends to remove the ($2J + 1$)-fold degeneracy associated with total angular momentum $J$ of the $4f$ ground state multiplet. The magnetic ordering in compounds having CEF-split nonmagnetic singlet ground state are of particular interest. The magnetic properties of such systems critically depend on the relative strength of the crystal electric field and the exchange field between the rare earth ions. The magnetic moment associated with the $4f$ electrons can be completely quenched if the former dominates over the latter \cite{Trammel1963}.  No magnetic ordering can occur in singlet ground state system unless the exchange interaction $\mathcal{J}_{\rm ex}$ exceeds a certain critical value relative to the crystal electric field splitting energy $\Delta$ between the ground state and the excited state coupled by the matrix element $\alpha$ \cite{Cooper1967}. The critical value of $\mathcal{J}_{\rm ex}$ for the case of singlet ground state and singlet first excited state is given by the condition $4 \mathcal{J}_{\rm ex} \alpha^2 / \Delta = 1$ \cite{Wang1968}. The system undergoes a self-induced spontaneous ordering above this critical value of $\mathcal{J}_{\rm ex}/\Delta$. Pr$_3$Tl, TmNi$_2$, Pr$_3$In and PrRu$_2$Si$_2$ are few of the well known CEF-split singlet ground state systems undergoing the self-induced spontaneous ordering \cite{Birgeneau1971,Andres1972,Deutz1986,Deutz1989,Goremychkin1989,Christianson2005,Christianson2007,Mulders1997}.

With a $J$ value of 4 for Pr$^{3+}$ ion, the Pr-based compounds offer an interesting platform to explore the singlet ground state system. The action of crystal electric field on the $2J + 1~(= 9) $-fold degenerate multiplet of Pr$^{3+}$ ion splits it in such a way that a CEF-split singlet state is always present for all type of crystallographic symmetry \cite{Walter1984}. Some of the singlet ground state Pr-based systems are known to exhibit exotic behavior, such as the observation of unconventional heavy-fermion superconductivity in PrOs$_{4}$Sb$_{12}$ \cite{Bauer, Goremychkin2004}. Heavy fermion behavior in PrOs$_{4}$Sb$_{12}$ is proposed to arise by an unconventional mechanism of the excitonic mass enhancement by low lying crystal field excitations. A similar mechanism is also believed to be the origin of the heavy fermion behavior in singlet ground state system PrRh$_{2}$B$_{2}$C \cite{Anand2009}.

In recent work on induced moment CEF singlet ground state system PrAu$_{2}$Si$_{2}$, an interesting aspect of competition between the crystal electric field and exchange field was noted. A systematic study of the electronic ground state in PrAu$_2$(Si$_{1-x}$Ge$_x$)$_2$ revealed that the CEF effect plays an important role in introducing a frustration to the magnetic ground state of the system \cite{Krimmel,Goremychkin2008}. A novel mechanism due to the dynamic fluctuations of the crystal field levels was proposed to be the origin of the spin-glass behavior in PrAu$_{2}$Si$_{2}$ \cite{Goremychkin2008}. Inelastic neutron scattering study revealed a CEF-split singlet ground state in PrAu$_{2}$Si$_{2}$ with an excited state doublet at 0.72~meV \cite{Goremychkin2008,Goremychkin2007}. The exchange interaction in PrAu$_{2}$Si$_{2}$ is found to be very close to the critical value required for the induced moment magnetic ordering. As such the dynamic fluctuations of CEF levels destabilize the induced moment magnetism in PrAu$_{2}$Si$_{2}$ leading to spin-glass behavior in this compound. This novel aspect of competition between the crystal field effect and exchange interaction brings a new perspective to the mechanism of spin-glass behavior. Similar frustrated magnetic ground states and spin-glass behaviors are also observed in CEF-split singlet ground state systems PrRuSi$_{3}$ and PrRhSn$_{3}$ \cite{Anand2011,Anand2012a}. Here we report another CEF-split singlet ground state system PrIrSi$_{3}$ that exhibits a possible magnetically ordered ground state despite the dominant CEF interaction and large splitting energy between the ground state and the first excited state.

The physical properties of PrIrSi$_{3}$ have been investigated by dc magnetic susceptibility $\chi$, isothermal magnetization $M$, thermo-remnant magnetization $M$, ac magnetic susceptibility $\chi_{\rm ac}$, specific heat $C_{\rm p}$ and electrical resistivity $\rho$ as function of temperature $T$ and magnetic field $H$. The magnetic ground state of PrIrSi$_{3}$ is further probed by muon spin-relaxation ($\mu$SR) and inelastic neutron scattering (INS) measurements. The $\chi(T)$ shows a sharp anomaly due to magnetic phase transition at $T_{tr} = 12.2$~K. However, the $C_{\rm p}(T)$ data exhibit only a weak anomaly near 12~K\@. The $\mu$SR finds possible magnetic ordering below 30~K, apparently due to the presence of short range magnetic correlations well above the magnetic phase transition temperature. A singlet ground state is inferred both from $C_{\rm p}(T)$  and INS data. The INS spectra show two sharp inelastic excitations at 8.5~meV and 18.5~meV which are well represented by a CEF model. The analysis of INS data reveal a very large CEF splitting energy of 92~K between the ground state singlet and the first excited state doublet. As such the criterion of induced moment magnetism is not easily fulfilled in PrIrSi$_{3}$ unless there is another kind of interaction in addition to the RKKY interaction. Thus, the observation of long range magnetic ordering in singlet ground state PrIrSi$_{3}$ is very surprising and may bring further insight to the interplay of crystal field effect and exchange interaction in singlet ground state system.

\section{\label{Sec:Exp} Experimental}

A polycrystalline sample of PrIrSi$_{3}$ was prepared by the standard arc melting technique. The stoichiometric mixture of high purity elements (Pr:  99.9\%, Ir: 99.99\%, Si 99.999\%) were arc melted on a water cooled copper hearth under the titanium gettered inert argon atmosphere. The sample was homogenized by flipping and re-melting several times. The arc melted buttons were wrapped in tantalum foil and annealed at 900 $^{\circ}$C for a week under the dynamic vacuum. The phase purity and crystal structure of the sample was determined by the powder x-ray diffraction (XRD) using Cu K$_{\alpha}$ radiation. The dc magnetization measurements were performed by using a commercial superconducting quantum interference device magnetometer (MPMS, Quantum Design Inc.) and the ac magnetic susceptibility measurements were performed by using a physical properties measurement system (PPMS, Quantum Design Inc.). The specific heat measurements were performed by the relaxation method using the PPMS. The electrical resistivity measurements were performed by the standard four probe ac technique using the PPMS.

The muon spin relaxation and inelastic neutron scattering measurements were performed at the ISIS facility of the Rutherford Appleton Laboratory, Didcot, U.K.  The $\mu$SR measurements were carried out on the MuSR spectrometer with the detectors in a longitudinal configuration. The powdered sample was mounted on a high purity silver plate using diluted GE varnish and covered with kapton film which was cooled down to 1.0~K in a standard He-4 cryostat with He-exchange gas. The INS measurements were carried out on the MARI time of flight spectrometer. The powdered sample was wrapped in a thin Al-foil and mounted inside a thin walled Al-can which was cooled down to 4.5 K inside a top-loading closed cycle refrigerator with He-exchange gas. The neutrons with incident energies $E_{i} = 6$, 20 and 40~meV were used and the scattered neutrons were detected between 3$^{\circ}$ and 135$^{\circ}$ scattering angle.

\section{\label{Sec:crystal} Crystallography}

\begin{figure}
\begin{center}
\includegraphics[width=8.5cm, keepaspectratio]{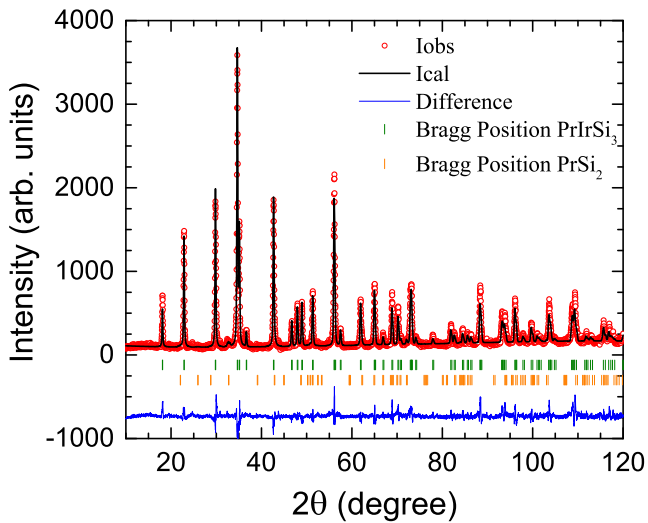}
\caption {(Color online) Powder x-ray diffraction pattern of PrIrSi$_{3}$ recorded at room temperature. The solid line through the experimental points is the two-phase Rietveld refinement profile calculated for BaNiSn$_{3}$-type tetragonal (space group $I4\,mm$) and ThSi$_2$-type tetragonal (space group $ I4_1/amd$) structures. The short vertical bars indicate the Bragg peak positions. The lowermost curve represents the difference between the experimental and calculated intensities.}
\label{fig:XRD}
\end{center}
\end{figure}

\begin{table}
\caption{\label{tab:XRD} Crystallographic parameters obtained from the structural Rietveld refinement of powder XRD data of PrIrSi$_{3}$. The refinement quality parameter $\chi ^2= 3.24$.}
\begin{ruledtabular}
\begin{tabular}{llccc}

\multicolumn{2}{l}{Structure} &\multicolumn{3}{l} {BaNiSn$_{3}$-type tetragonal} \\
\multicolumn{2}{l}{Space group} & \multicolumn{2}{l} {$I4\,mm$ (No. 107)}\\
\multicolumn{2}{l}{Formula units/unit cell ($Z$)}  &  2 \\
\multicolumn{2}{l}{\underline{Lattice  parameters}} \\
\multicolumn{2}{l}{\hspace{0.8cm} $a$ ({\AA})}            			&  4.2367(2)  \\	
\multicolumn{2}{l}{\hspace{0.8cm} $c$ ({\AA})}          			&  9.8050(4)  \\
\multicolumn{2}{l}{\hspace{0.8cm} $V_{\rm cell}$  ({\AA}$^{3}$)} 	&  176.00(1)  \\
\\
\multicolumn{2}{l}{\underline{Atomic coordinates}} \\
\hspace{0.7cm}Atom &\hspace{0.5cm} Wyckoff & $x$ & $y$ & $z$  \\
& \hspace{0.5cm} position \\
\hspace{1cm}Pr & \hspace{1cm} 2a & 0 & 0 & 0 \\
\hspace{1cm}Ir & \hspace{1cm} 2a & 0 & 0 & 0.6568(3) \\
\hspace{1cm}Si1 & \hspace{1cm} 2a & 0 & 0 & 0.412(2) \\
\hspace{1cm}Si2 & \hspace{1cm} 4b & 0 & 1/2 & 0.2582(9)\\
\end{tabular}
\end{ruledtabular}
\end{table}

Figure~\ref{fig:XRD} shows the x-ray diffraction pattern of PrIrSi$_{3}$ which was collected on the powdered sample at room temperature. The XRD data were analyzed by Rietveld structural refinement using the software {\tt FullProf} \cite{Rodriguez1993} which revealed a BaNiSn$_{3}$-type body-centered tetragonal structure (space group $I4\,mm$) for PrIrSi$_{3}$.  The Rietveld refinement profile for this structure is shown in figure~\ref{fig:XRD}. The crystallographic parameters obtained from the refinements are listed in Table~\ref{tab:XRD}. The single phase nature of the sample is evident from the refinement. A few weak impurity peaks are also present most of which could be accounted by two phase refinement, the impurity phase being PrSi$_2$ (1.35 wt\%). During refinement the occupancies were kept fixed to unity and the thermal parameters $B$ were set to zero. No significant improvement in the refinement quality parameters or in the $z$-positions of atoms was obtained upon allowing variations in the occupancies or $B$.

\begin{figure}
\begin{center}
\includegraphics[width=3in, keepaspectratio]{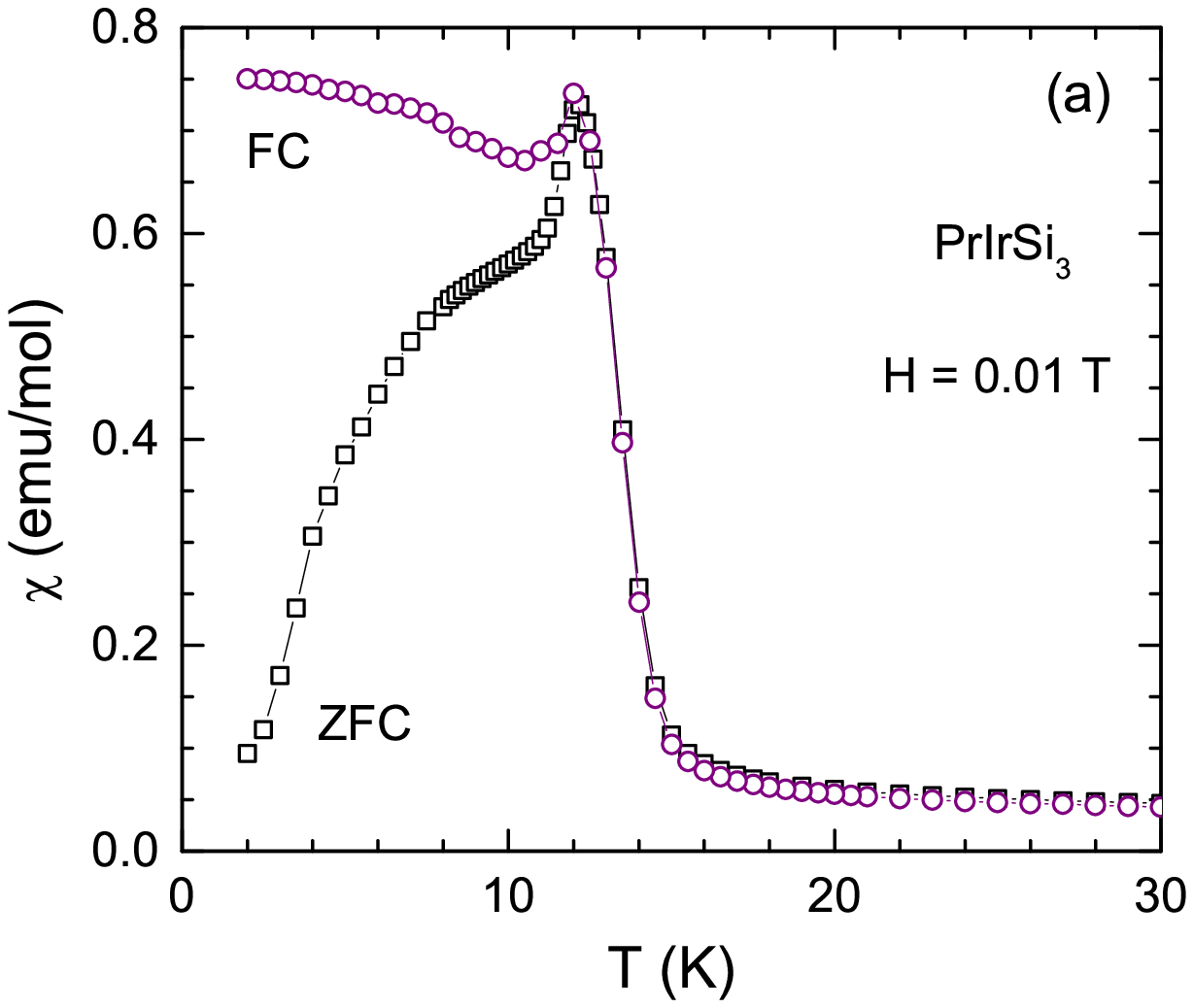} \vspace{0.1in}
\includegraphics[width=3in, keepaspectratio]{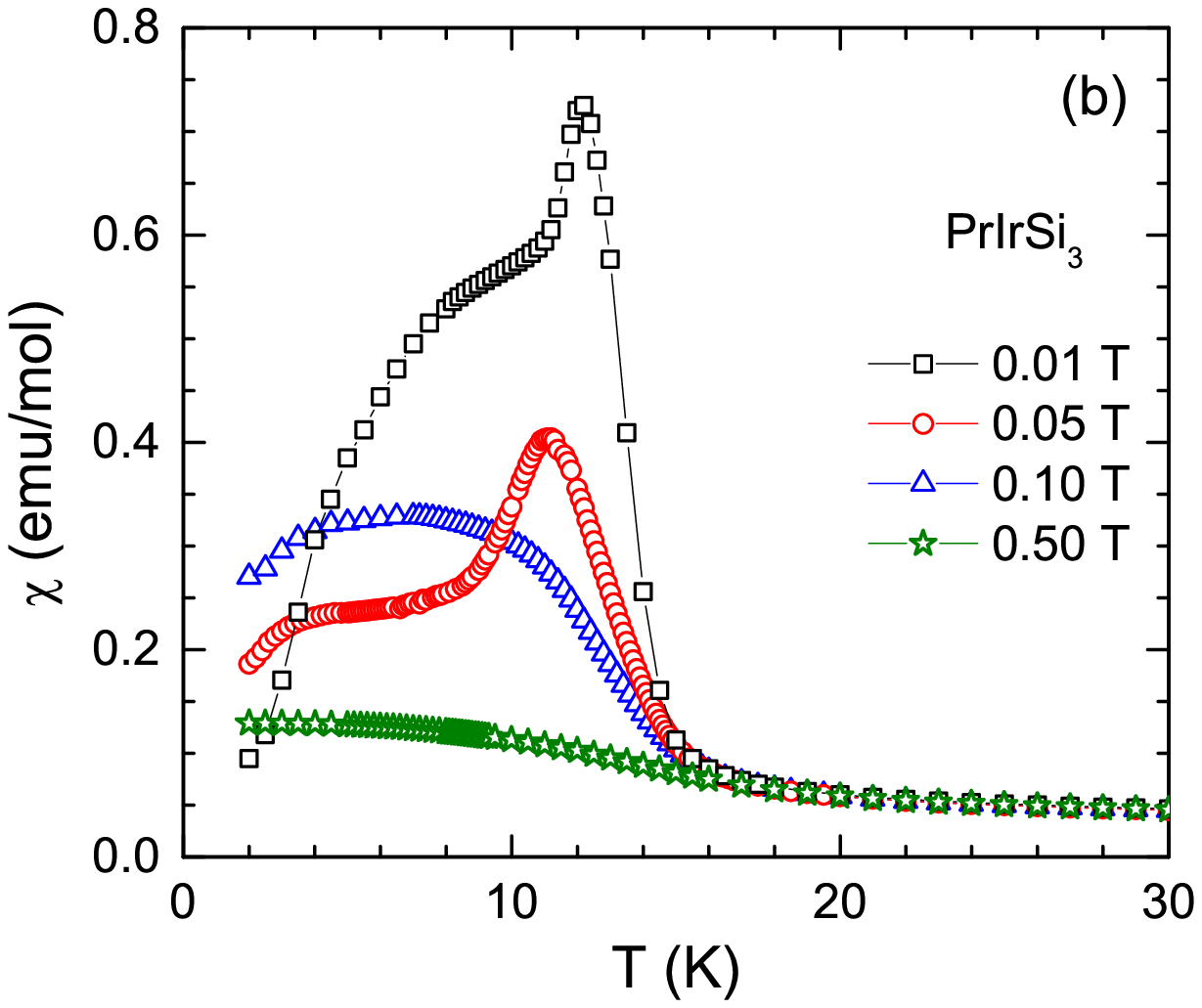}
\caption {(Color online) (a) The temperature $T$ dependence of zero field cooled (ZFC) and field cooled (FC) dc magnetic susceptibility $\chi$ of PrIrSi$_{3}$ in the temperature range 2 -- 30 K measured in a magnetic field $H =0.01$~T\@. (b) The ZFC $\chi(T)$ data measured at different $H$.}
\label{fig:Chi-LowH}
\end{center}
\end{figure}

The impurity phase PrSi$_2$ is reported to order ferromagnetically below 11~K followed by another transition at 7~K \cite{Pierre1988a,Pierre1988b,Lebaroo1989,Dhar1994}. The dc magnetic susceptibility of PrIrSi$_{3}$ exhibits sharp anomaly at 12.2~K and a broad hump at 7.5~K (figure~\ref{fig:Chi-LowH}) which are very close to the transition temperatures of PrSi$_2$. Therefore, anomalies observed in magnetic data of PrIrSi$_{3}$ may at first glance be naively connected with the traces of PrSi$_2$. However, a closer look at the temperature dependence of the ac magnetic susceptibility, resistivity and specific heat of PrSi$_2$ and PrIrSi$_{3}$ in the vicinity of transition show different behavior and the distinctness of magnetic behavior of PrIrSi$_{3}$ is evident. The ac susceptibility of PrSi$_2$ shows two well pronounced peaks at 11~K and 7~K \cite{Lebaroo1989}. In contrast the ac susceptibility of PrIrSi$_{3}$ shows sharp peak at 12.2~K and a kink at 13.5~K, and there is no peak near 7~K\@. The isothermal magnetization of PrSi$_2$ exhibits a rapid increase up to 1.5~T whereas the isothermal magnetization of PrIrSi$_{3}$ exhibits a rapid increase only up to 0.2~T\@. Further, PrSi$_2$ which forms in ThSi$_2$-type tetragonal (space group {\it I4$_1$/amd}) at room temperature undergoes a structural phase transition to GdSi$_2$-type orthorhombic structure (space group {\it Imma}) below 180~K and a well defined anomaly due to structural transition is observed in resistivity of PrSi$_2$ \cite{Pierre1988a,Pierre1988b}. However, no such anomaly is observed in our resistivity of PrIrSi$_{3}$ (see figure~\ref{fig:rho}). Furthermore in our inelastic neutron scattering spectra we do not see the reported \cite{Pierre1988b} excitations near 9.7~meV, 5.5~meV and 2~meV which would otherwise be due to the presence of PrSi$_2$, suggesting that the contribution from the impurity phase is of no consequence.  All of these lead to the compelling conclusion that our observed magnetic behavior is intrinsic to the title compound PrIrSi$_{3}$. We therefore believe that the presence of tiny amount of impurity does not affect the results and conclusions reached on the bulk properties of PrIrSi$_{3}$ presented here. The reproducibility of anomalies in the magnetic data have been checked on two different batches of PrIrSi$_{3}$.

\begin{figure}
\begin{center}
\includegraphics[width=3.2in, keepaspectratio]{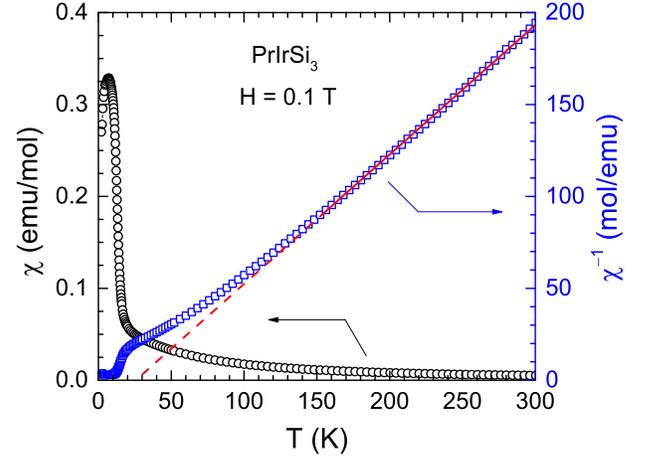}
\caption {(Color online) The temperature $T$ dependence of zero field cooled dc magnetic susceptibility $\chi$ of PrIrSi$_{3}$ and its inverse $\chi^{-1}$ in the temperature range 2--300 K measured in a magnetic field $H =0.1$~T\@. The solid line represents the fit to Curie-Weiss law for $150~{\rm K} \leq T \leq 300$~K and the dashed line is an extrapolation.}
\label{fig:Chi}
\end{center}
\end{figure}

\section{\label{Sec:Magnetization} Magnetic Susceptibility and Magnetization}

Figures~\ref{fig:Chi-LowH} and \ref{fig:Chi} show the temperature dependence of zero field cooled (ZFC) and field cooled (FC) dc magnetic susceptibility $\chi$ of PrIrSi$_{3}$ measured at different applied magnetic field $H$. A magnetic phase transition at temperature $T_{tr} = 12.2$~K is inferred from the low-$T$ $\chi(T)$ data. At low $H$, e.g. at $H = 0.01$~T, the low-$T$ $\chi(T)$ [figure~\ref{fig:Chi-LowH}(a)] exhibits a rapid increase below 15~K and peak at 12.2~K with a broad hump near 7.5~K\@. An irreversibility is also observed between the ZFC and FC $\chi(T)$ data below 12.2~K\@. Such a behavior of $\chi(T)$ reflects the presence of ferromagnetic correlation. An increase in $H$ shifts the $\chi(T)$ peak position towards the lower temperature, e.g., at $H = 0.05$~T the 12.2~K peak shifts to 11.1~K and 7.5~K anomaly to $\sim 4$~K, and reduces the magnitude of $\chi$ [figure~\ref{fig:Chi-LowH}(b)]. At $H = 0.1$~T both the anomalies merge together and appear as a very broad anomaly centered around 7~K\@.

At high temperatures the $\chi(T)$ data follow the Curie-Weiss behavior, $\chi(T) = C/(T-\theta_{\rm p})$. From a linear fit of the inverse magnetic susceptibility $\chi^{-1}(T)$ data measured at $H=0.1$~T in the temperature range 150~K~$\leq T\leq$~300~K (the solid line in figure~\ref{fig:Chi}) we obtain the Curie constant $C = 1.42(1)$~emu\,K/mol and paramagnetic Weiss temperature $\theta_{\rm p} = +26(2)$~K. The effective moment $\mu_{\rm eff}$ obtained from $C$ using the relation $\mu_{\rm eff} \approx \surd (8C) = 3.37(1)\, \mu_{\rm B}$. The effective moment is slightly smaller than the theoretically expected value of effective moment for Pr$^{3+}$ ions ($3.58 \, \mu_{\rm B}$). The positive $\theta_{\rm p}$ reflects a dominant ferromagnetic interaction.

\begin{figure}
\begin{center}
\includegraphics[width=3in, keepaspectratio]{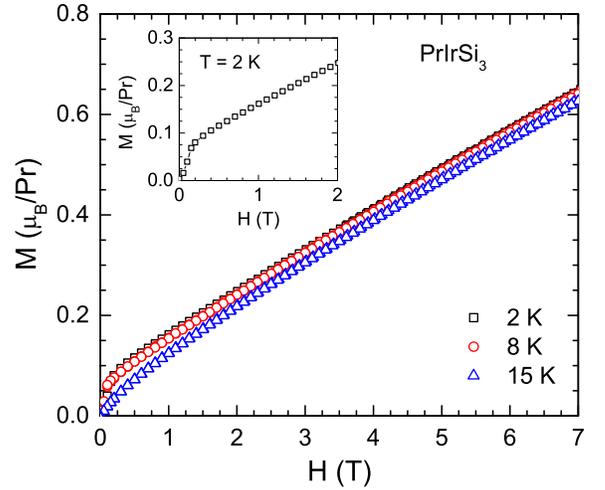}
\caption {(Color online) The dc isothermal magnetization $M$ of PrIrSi$_{3}$ as a function of magnetic field $H$ measured at the selected temperatures of 2, 8, and 15~K. Inset: An expanded plot of $M(H)$ isotherm at 2~K for $H \leq 2$~T\@.}
\label{fig:MH}
\end{center}
\end{figure}

Figure~\ref{fig:MH} shows the $H$ dependence of isothermal magnetization $M$ of PrIrSi$_{3}$ measured at selected temperatures of 2, 8, and 15~K\@. At low-$T$ (e.g., at 2~K and 8~K) the $M(H)$ isotherm exhibits ferromagnetic-like spontaneous magnetization.  It is seen that at 2~K the $M$ increases very rapidly initially for $H \leq 0.2$~T and then the increase rate gradually decreases and eventually for $H \geq 0.6$~T the $M(H)$ isotherm is almost linear in $H$\@. The $M(H)$ isotherm at 15~K is almost linear in $H$ for $H > 1.0$~T\@. The nonlinearity for $H\leq 1.0$~T can be due to the presence of short range correlations above $T_{tr}$ and/or crystal field effect. Further we see that at 2~K and 7~T, the magnetization attains a value of $\sim  0.65\, \mu_{\rm B}$, which is very low compared to the theoretical value of saturation magnetization $M_{\rm s}$ = 3.2\,$\mu_{\rm B}$ for Pr$^{3+}$ ions. This reduction in $M$ can be attributed to the crystal field effect inferred from the specific heat and resistivity data in next section.

In the absence of CEF, the magnetization is given by \cite{Blundell2001}
\begin{equation}
\label{eq:M1}
M = M_{\rm s} B_J(x)
\end{equation}
where $B_J(x)$ is the Brillouin function given by
\begin{equation}
\label{eq:M2}
B_J(x)=\frac{2J+1}{2J}\coth\left(\frac{2J+1}{2J}x\right)-\frac{1}{2J}\coth\left(\frac{x}{2J}\right)
\end{equation}
with
\begin{equation}
x = \frac{g_J \mu_{\rm B} J H}{k_{\rm B}T}.
\end{equation}
For Pr$^{3+}$ Land\'e $g$ factor $g_J =4/5$. From Eqs.~(\ref{eq:M1}) and (\ref{eq:M2}) for $T=2$~K and $H=7$~T we obtain $M=3.06\,\mu_{\rm B}$ which is much higher than the observed $M\approx 0.65\,\mu_{\rm B}$. This suggests that the CEF is responsible for the reduced $M$ in PrIrSi$_{3}$. A reduced value of $M\approx 0.9\,\mu_{\rm B}$ was also observed in isostructural PrRuSi$_{3}$ \cite{Anand2011} and $M\approx 0.6\,\mu_{\rm B}$ is observed in isostructural PrRhSi$_{3}$ at $T=2$~K and $H=7$~T \cite{Anand2013}. On the other hand for isostructural PrCuAl$_{3}$ $M\approx 2.05\,\mu_{\rm B}$ at $T=2$~K and $H=7$~T \cite{Adroja2012} and that for cubic PrRhSn$_{3}$ is $M\approx 1.25\,\mu_{\rm B}$ \cite{Anand2012a}.

\begin{figure}
\begin{center}
\includegraphics[width=3in, keepaspectratio]{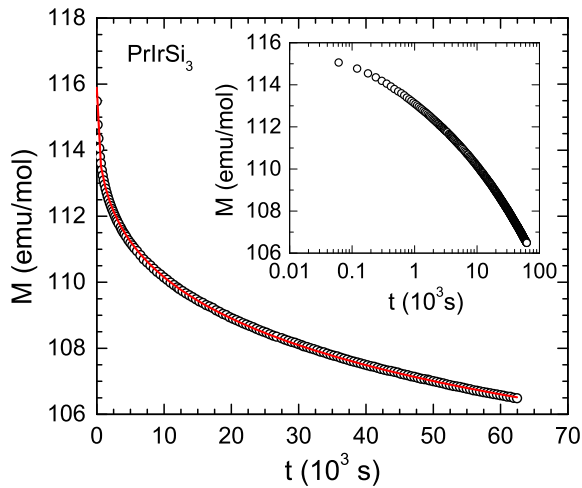}
\caption {(Color online) The time $t$ dependence of thermo-remnant magnetization $M$ of PrIrSi$_{3}$ measured at 2~K after switching off the cooling magnetic field of 0.05~T\@. The solid curve is the fit to the superposition of a stretched exponential and a constant term. Inset: $M(t)$ data plotted on a semi-log scale.}
\label{fig:TRM}
\end{center}
\end{figure}

Figure~\ref{fig:TRM} shows the time $t$ dependence of thermo-remnant magnetization $M$ of PrIrSi$_{3}$. The time dependence of $M$ was recorded by field cooling the sample in a magnetic field of 0.05~T from 50~K to 2~K, and then the field cooled isothermal remnant magnetization was measured at 2~K after switching off the magnetic field. As seen from figure~\ref{fig:TRM} the $M(t)$ decays very slowly with time and remains nonzero even after 62000~s\@. A semi-logarithmic plot of $M(t)$ data is shown in the inset of figure~\ref{fig:TRM} which clearly shows that $M(t)$ data do not follow the logarithmic decay, $M(t) = M_{0}- S \log(t)$. However, the $M(t)$ data could be well fitted by a superposition of a stretched exponential and a constant term, \cite{Sinha,Taniyama}
\begin{equation}
M(t) = M_{0} + M_{1}\, exp\left[-\left(\frac{t}{\tau}\right)^{1-n}\right]
\label{eq:TRM}
\end{equation}
 where $\tau$ is the mean relaxation time. The constant term $M_{0}$ is expressed as the longitudinal spontaneous magnetization coexisting with the frozen transverse spin component \cite{Gabay}. A fit of $M(t)$ data by Eq.~(\ref{eq:TRM}) is shown by the solid curve in figure~\ref{fig:TRM} which is obtained for $M_{0} = 98.6(5)$~emu/mole, $M_{1} = 17.3(5)$~emu/mole, $\tau = 123761(176)$~s and $n = 0.64(1)$. A similar analysis of $M(t)$ for spin-glass system PrRuSi$_{3}$ have yielded $M_{0} = 53.0$~emu/mole, $M_{1} = 2.5$~emu/mole, $\tau = 13357$~s and $n = 0.62$ \cite{Anand2011}. The mean relaxation time $\tau$ of PrIrSi$_{3}$ is extremely large and reflect a very slow decay of $M$\@.

\begin{figure}
\begin{center}
\includegraphics[width=3in, keepaspectratio]{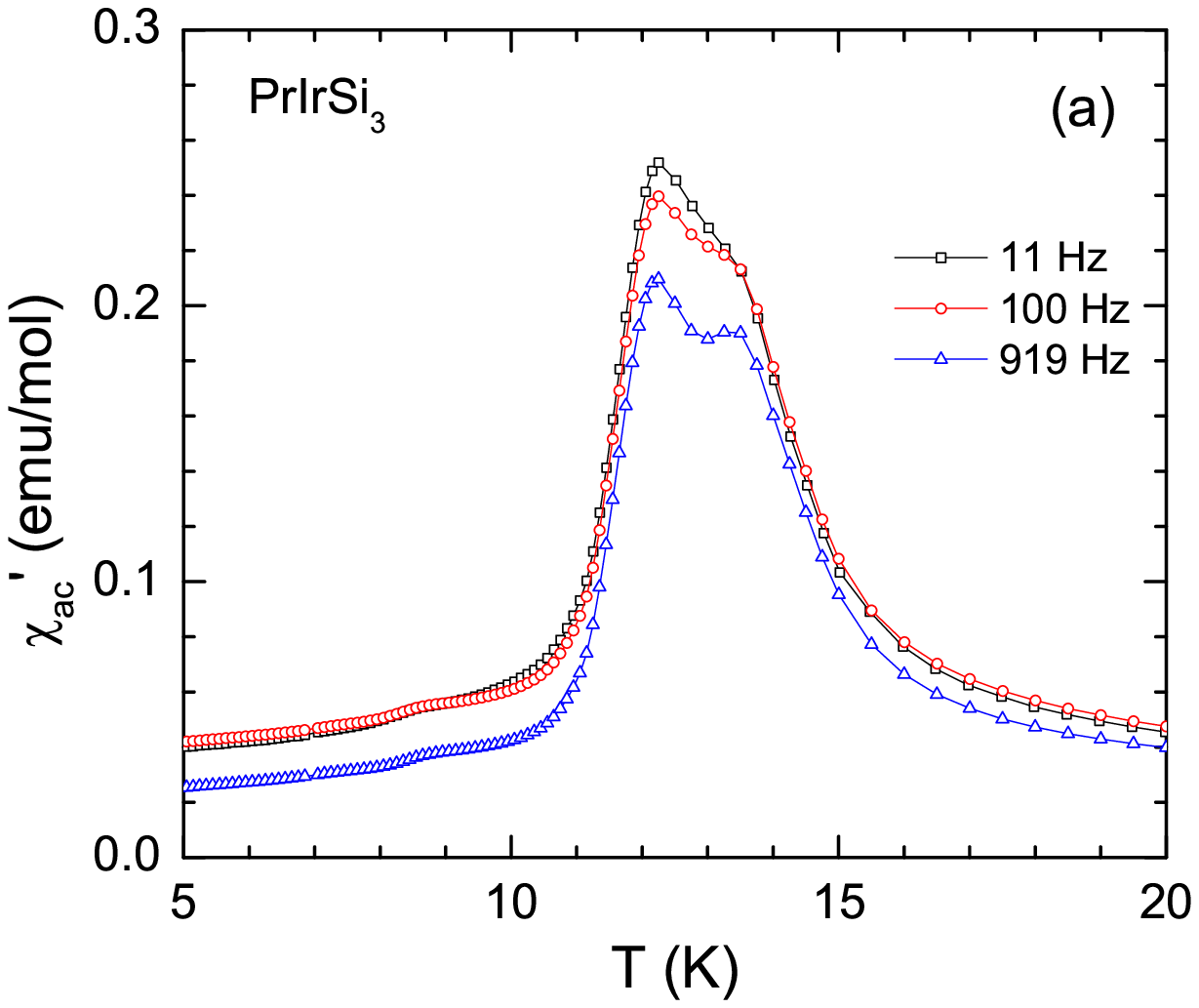} \vspace{0.1in}
\includegraphics[width=3in, keepaspectratio]{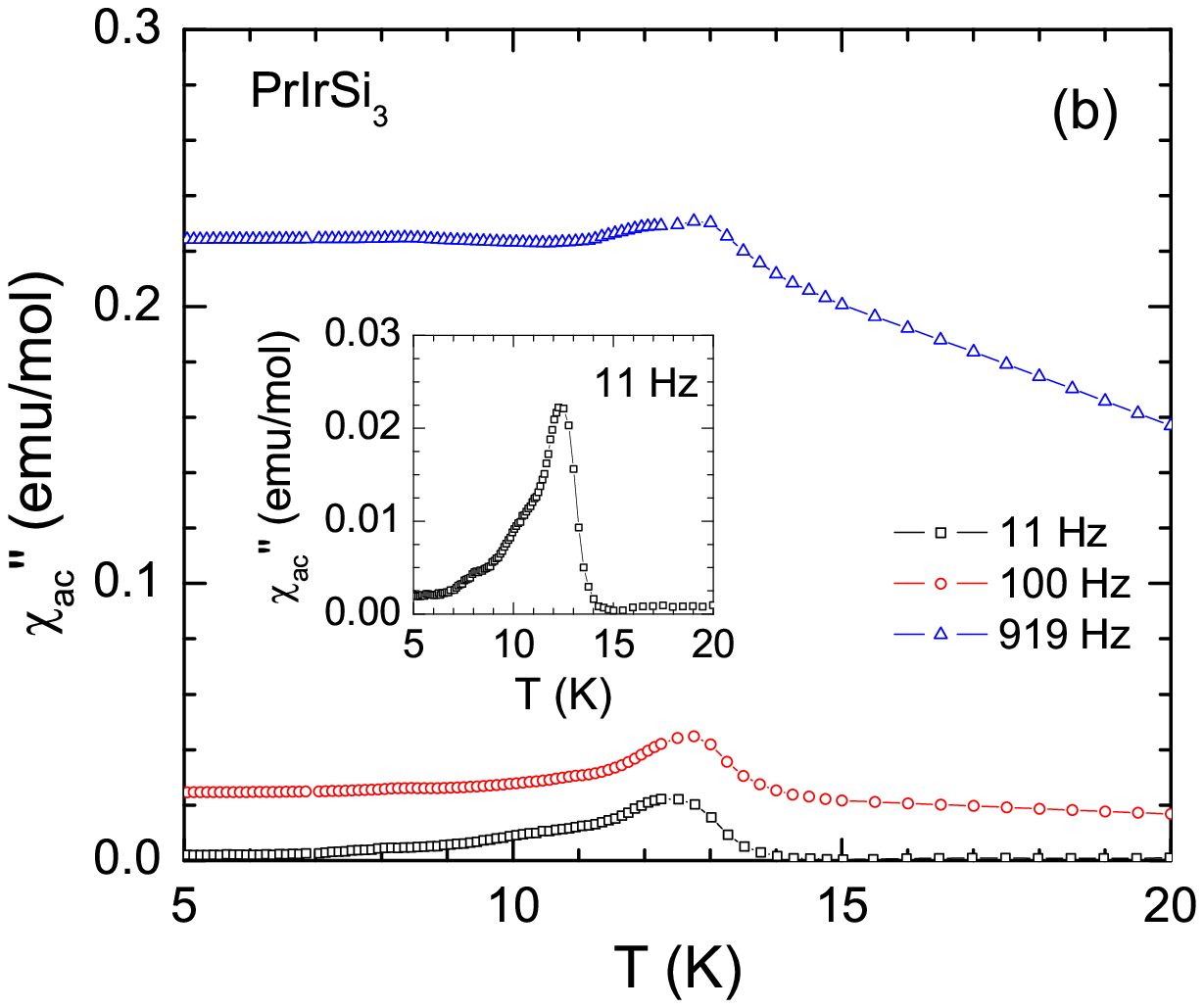}
\caption {(Color online) The temperature $T$ dependence of (a) real $\chi_{\rm ac}'$ and (b) imaginary $\chi_{\rm ac}''$ parts of the ac magnetic susceptibility $\chi_{\rm ac}$ of PrIrSi$_{3}$ measured at different frequencies from 11~Hz to 919~Hz in an applied ac magnetic field of 1.5~mT\@. Inset in (b): An expanded plot of $\chi_{\rm ac}''(T)$ at 11 Hz.}
\label{fig:Chiac}
\end{center}
\end{figure}

Figure~\ref{fig:Chiac} shows the $T$ dependence of the real $\chi_{\rm ac}'$ and imaginary $\chi_{\rm ac}''$ parts of ac susceptibility $\chi_{\rm ac}$ data of PrIrSi$_{3}$ measured at three different frequencies $\nu$ under an applied ac magnetic field of 1.5~mT. As seen from figure~\ref{fig:Chiac}(a) a sharp anomaly at 12.2~K is observed in $\chi_{\rm ac}'(T)$ at 11~Hz together with a kink at 13.5~K and a hardly detectable hump near 7.5~K\@. No noticeable change is observed in the anomaly temperature with an increase in $\nu$ for $\nu \leq 5.5$~kHz except that the 13.5~K anomaly becomes more pronounced with increasing frequency. A weak temperature shift in both 13.5~K and 12.2~K anomalies towards the lower temperature is observed with the increase in $\nu$ for $5.5~{\rm kHz} \leq \nu \leq 10$~kHz (data not shown). The magnitude of $\chi_{\rm ac}'$ decreases with increasing frequency and becomes negative for $\nu \geq 2700$~Hz (data not shown). Such a behavior of $\chi_{\rm ac}'$ with increasing $\nu$ can be attributed to skin effect due to the confinement of ac magnetic flux in a surface layer which is possible because of the high conductivity of the sample at low temperature ($\rho_{0} \approx 1~\mu\Omega$\,cm). For a metal the skin depth $\delta \sim \surd (\rho / \nu) $, \cite{Jackson} therefore at high frequency the ac flux is confined to a skin layer only, leading to the observed diamagnetic behavior in $\chi_{\rm ac}'$ at higher frequencies.

On the other hand, $\chi_{\rm ac}''(T)$ exhibits sharp anomaly near 12.2~K at 11~Hz, however, there is no anomaly near 13.5~K corresponding to that observed in $\chi_{\rm ac}'$. The hump near 7.5~K is very weak. At increased frequency an anomaly appears at 13~K for $\nu \geq 919$~Hz. The magnitude of $\chi_{\rm ac}''$ increases very rapidly with increase in frequency. For $\chi_{\rm ac}''$ too a weak temperature shift in both 12.2~K and 13~K anomalies towards the lower temperature is observed with the increase in $\nu$ for $5.5~{\rm kHz} \leq \nu \leq 10$~kHz (data not shown).

\begin{figure}
\begin{center}
\includegraphics[width=3in]{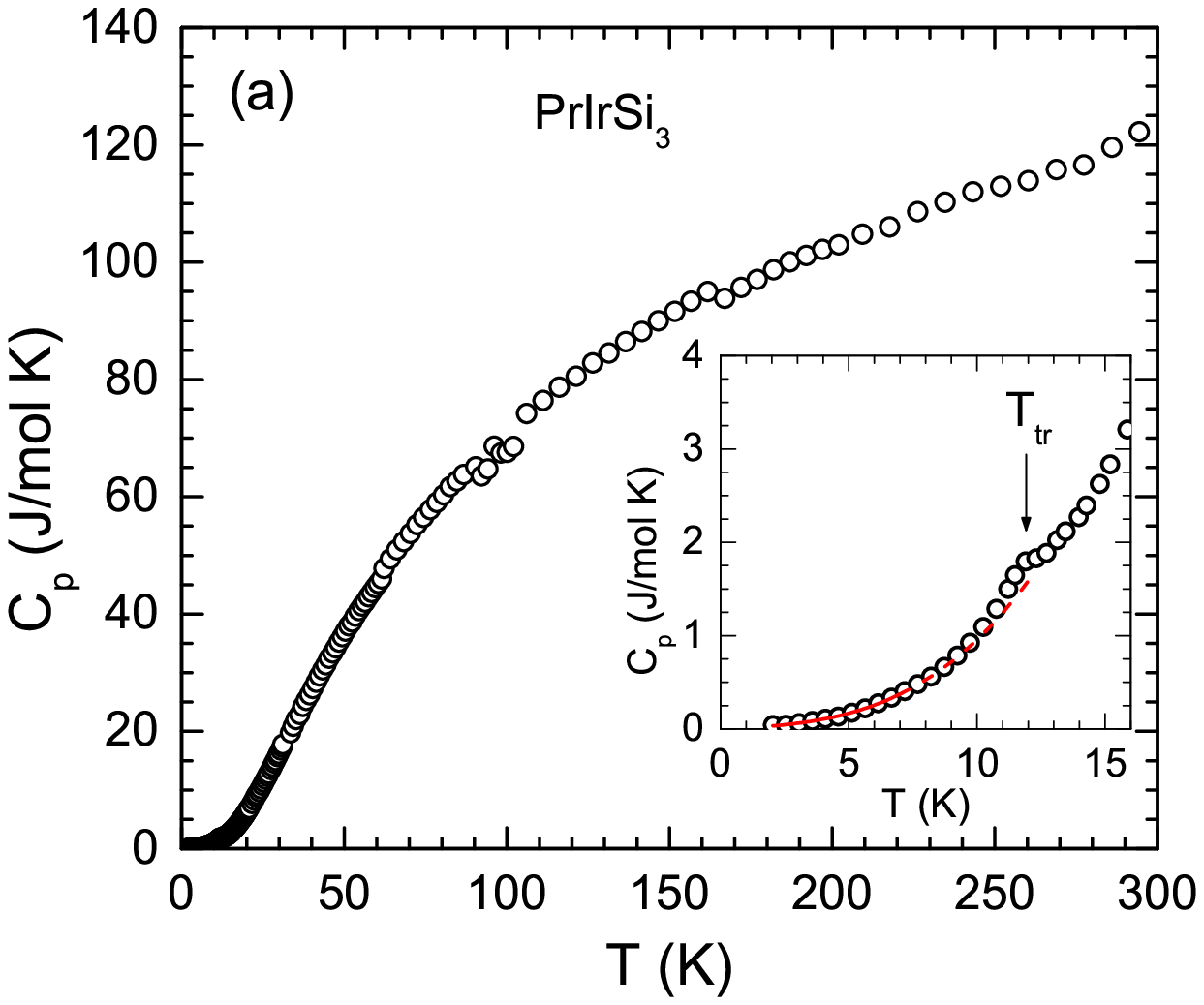} \vspace{0.1in}
\includegraphics[width=3in]{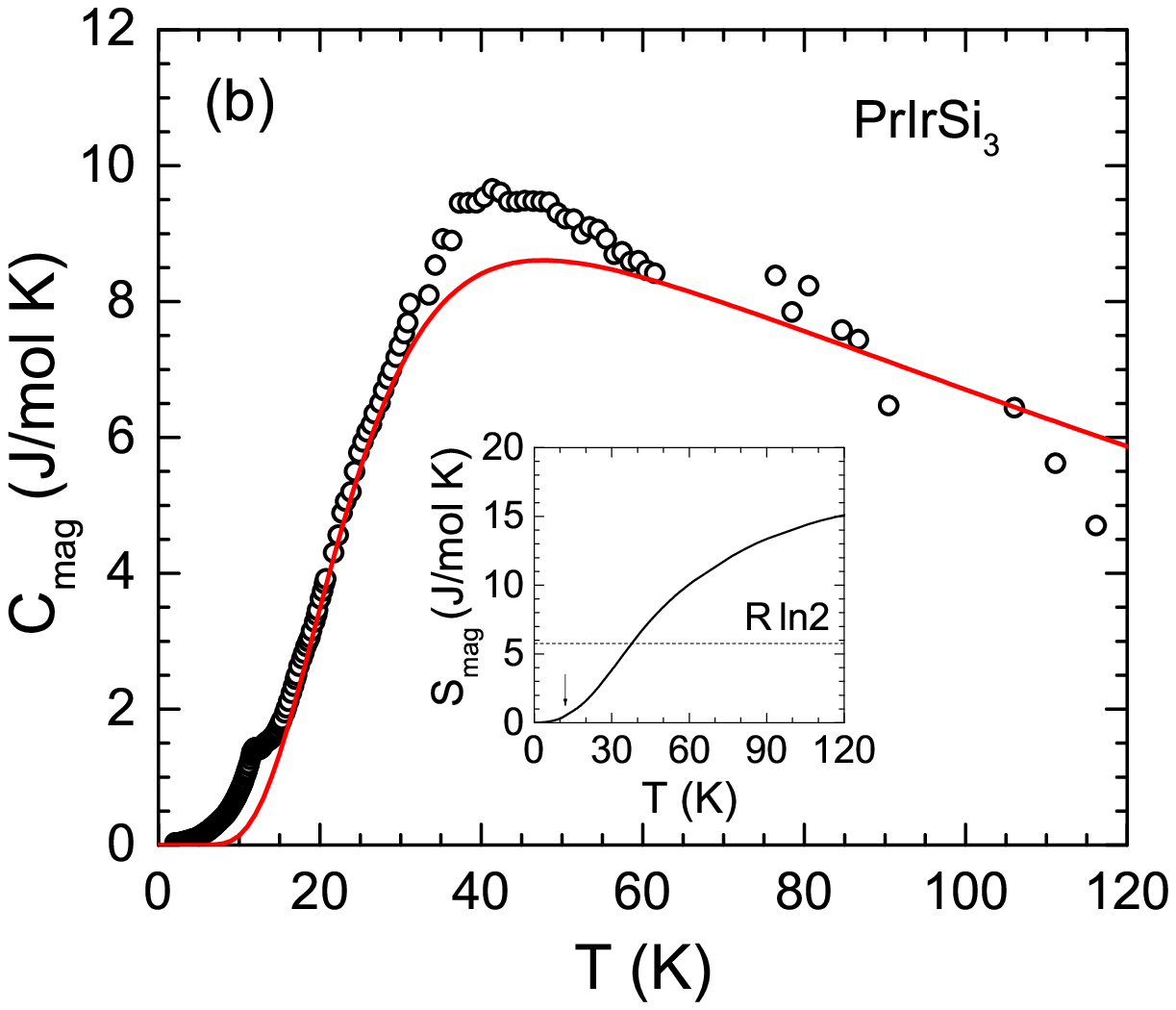}
\caption {(Color online) (a) The temperature $T$ dependence of specific heat $C_{\rm p}$ of PrIrSi$_{3}$ measured in zero magnetic field in the temperature range 2--300~K\@. Inset: An expanded view of low-$T$ $C_{\rm p}(T)$ data below 16~K\@. The solid curve in the inset is the fit by $C_{\rm p}(T) = \gamma T +\beta T^{3}$ below 7~K and the dashed line is an extrapolation. (b) The magnetic contribution to specific heat $C_{\rm mag} (T)$. The solid curve represents the crystal electric field contribution to specific heat according to the crystal field level scheme deduced from the inelastic neutron scattering data. Inset: Magnetic contribution to entropy $S_{\rm mag}(T) $. The arrow marks the transition temperature.}
\label{fig:HC}
\end{center}
\end{figure}

\section{\label{Sec:HC_Rho} Specific Heat and Electrical Resistivity}

Figure~\ref{fig:HC} shows the $T$ dependence of the specific heat $C_{\rm p}$ data of PrIrSi$_{3}$ measured at constant pressure p\@. As shown in the inset of figure~\ref{fig:HC}(a), the $C_{\rm p}(T)$ data present a weak anomaly near 12~K. The low-$T$ $C_{\rm p}(T)$ data follow $C_{\rm p}(T) = \gamma T +\beta T^{3}$, where $\gamma$ and $\beta$ are coefficients to the electronic and lattice contributions to specific heat, respectively. On fitting the $C_{\rm p}(T)$ data of PrIrSi$_{3}$ by this relation in $2~{\rm K} \leq T \leq 7$~K we obtain the Sommerfeld coefficient $\gamma = 12.5(6)$~mJ/mole\,K$^{2}$ and $\beta=0.83(2)$~mJ/mole\,K$^{4}$. The fit is shown in the inset of figure~\ref{fig:HC}(a). The Debye temperature $\Theta_{\rm D}$ estimated from $\beta$ using the relation $\Theta_{\rm D} = (12 \pi^{4} p R /{5 \beta} )^{1/3}$, where $R$ is the molar gas constant and $p=5$ is the number of atoms per formula unit, is $\Theta_{\rm D}$ = 227(1)~K\@.

The magnetic contribution to specific heat $C_{\rm mag}(T)$ is shown in figure~\ref{fig:HC}(b). The $C_{\rm mag}(T)$ was estimated by subtracting off the lattice contribution equal to the specific heat of isostructural LaOsSi$_{3}$ \cite{Xu2013}. The effect of crystal electric field is reflected as a broad Schottky-type anomaly centered around 40~K in $C_{\rm mag}(T)$. The solid curve in figure~\ref{fig:HC}(b) represents the crystal field contribution to specific heat according to the CEF level scheme obtained from the analysis of inelastic neutron scattering data in Sec.~\ref{Sec:INS}. The magnetic contribution to entropy $S_{\rm mag}(T)$ was obtained by integrating the $C_{\rm mag}(T)/T$ versus $T$ plot and is shown in the inset of figure~\ref{fig:HC}(b). An extremely small value of $S_{\rm mag}$ at $T_{tr}$ which is only $\sim 9$\,\% of $R\,ln2$ suggests a CEF-split singlet ground state in PrIrSi$_{3}$ and is confirmed by the INS data (Section.~\ref{Sec:INS}).

\begin{figure}
\begin{center}
\includegraphics[width=8.5cm]{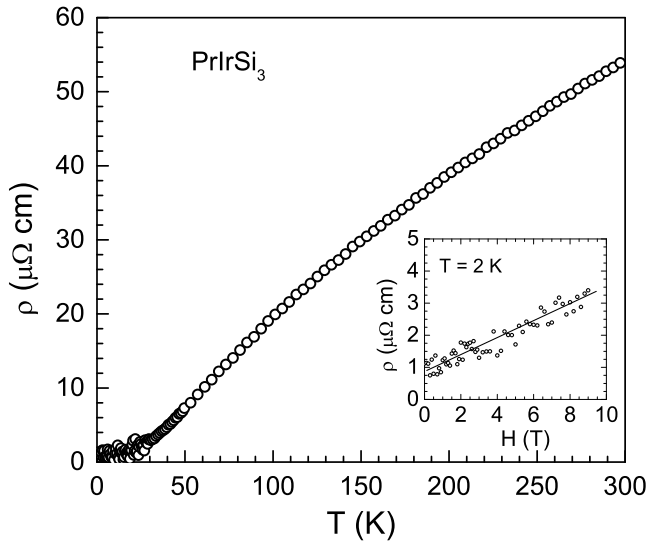}
\caption{(Color online) The temperature $T$ dependence of electrical resistivity $\rho$ of PrIrSi$_{3}$ measured in zero magnetic field. Inset: $\rho$ as a function of magnetic field $H$ at 2~K\@. The solid line is a guide to eye.}
\label{fig:rho}
\end{center}
\end{figure}

The $T$ dependence of the electrical resistivity $\rho$ of PrIrSi$_{3}$ is presented in figure~\ref{fig:rho}. The low value of residual resistivity $\rho_{0}\approx 1~\mu\Omega$\,cm at 2~K, and a high value of residual resistivity ratio $\rho_{\rm 300\,K}/\rho_{\rm 2\,K} \approx 50$ reveal the good quality of our sample. A metallic character is evident from the $T$ dependence of $\rho$. The effect of crystal electric field is reflected by a broad curvature in $\rho(T)$. The $H$ dependence of $\rho$ at 2~K is shown in the inset of figure.~\ref{fig:rho}. Though noisy, the basic trend of $\rho(H)$ is quite clear from the data, revealing that the $\rho$ increases linearly with increasing $H$. The $\rho$ is found to increase from its value of $\approx 1~\mu\Omega$\,cm at $H=0$~T to $\approx 3~\mu\Omega$\,cm at $H=9$~T thus revealing a large positive magnetoresistance (MR) of about 200\,\% at 2~K at 9 T\@.

\section{\label{Sec:muSR} Muon spin relaxation}

\begin{figure}
\begin{center}
\includegraphics[width=3in]{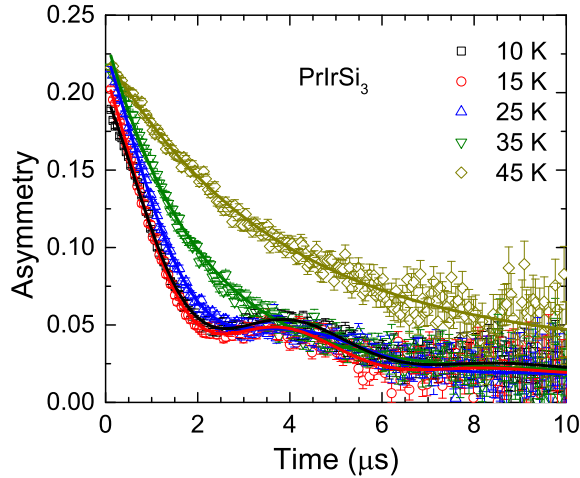}
\caption{\label{fig:MuSR1} (Color online) The zero field $\mu$SR spectra of PrIrSi$_{3}$ for few representative temperatures. The solid curves represent the fits to the data by the function described in the text.}
\end{center}
\end{figure}

\begin{figure}
\begin{center}
\includegraphics[width=3.3in]{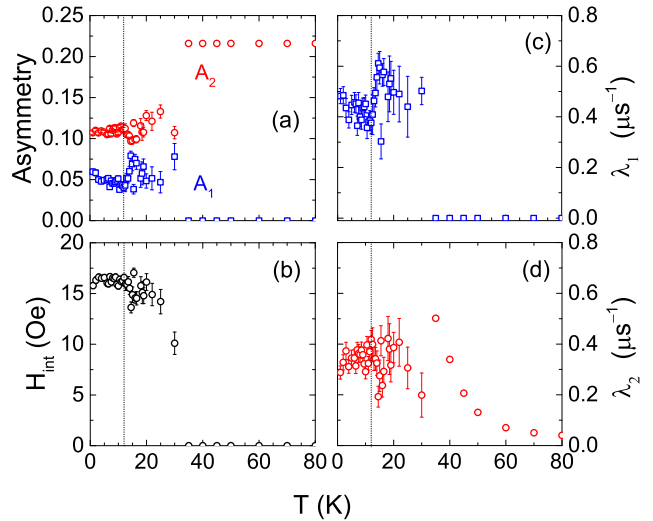}
\caption{\label{fig:MuSR2} (Color online) The temperature $T$ dependence of (a) the initial asymmetries $A_1$ and $A_2$, (b) the internal field $H_{\rm int}$, (c) the depolarization rate $\lambda_1$, and (d) the depolarization rate $\lambda_2$ obtained from the analysis of zero field $\mu$SR spectra of PrIrSi$_{3}$ collected at various temperatures. The vertical dotted lines mark $T_{tr}$ observed in the magnetic susceptibility in figure~\ref{fig:Chi-LowH}.}
\end{center}
\end{figure}

Figure~\ref{fig:MuSR1} shows the $\mu$SR spectra of PrIrSi$_{3}$ at few representative temperatures collected in zero field while warming the sample. At $T > 30$~K the $\mu$SR spectra are well described by a simple exponential (also called Lorentzian) decay function as the muons sense only paramagnetic fluctuations at these temperatures. However, for $T \leq 30$~K the $\mu$SR spectra are best described by an oscillating function with a Lorentzian envelope
\begin{equation}
 G_z(t)=A_1 \cos(\omega t + \varphi) \,e^{-\lambda_1 t} + A_2 \,e^{-\lambda_2 t}+ BG,
\end{equation}
where $A_{1}$ and $A_{2}$ are the initial asymmetries, $\omega = \gamma_{\mu} H_{\rm int}$ is the precession frequency ($\gamma_{\mu}$ is the gyromagnetic ratio of the muon and $ H_{\rm int}$ is the internal field at the muon site), $\varphi$ is the phase, $\lambda_1$ and $\lambda_2$ are the depolarization rates, and BG is a constant background. The phase angle accounts for the small misalignment of incident muon spins with respect to incident direction and small effect of zero time correction.

The fits of $\mu$SR data by this decay function are shown by the solid curves in figure~\ref{fig:MuSR1}. The parameters obtained from the fit of $\mu$SR data are plotted in figure~\ref{fig:MuSR2} as a function of temperature. The $T$ dependence of the asymmetries in figure~\ref{fig:MuSR2}(a) show a clear gain in asymmetry $A_{1}$ and loss of asymmetry A$_{2}$ at $T \leq 30$~K which is accompanied by an increased frequency and hence the internal field at the muon site as shown in figure~\ref{fig:MuSR2}(b). The temperature dependences of the muon depolarization rates $\lambda_1$ and $\lambda_2$ are shown in figures~\ref{fig:MuSR2}(c) and (d). An increase is observed in $\lambda_1$ at $T \leq 30$~K again indicating that the muons sense a possible magnetic phase transition near 30~K\@. The appearance of internal field at $T \leq 30$~K indicates that the muons sense a magnetically ordered state below 30~K\@. However, the observation of magnetically ordered state by muons at $T \approx 30$~K is much higher than the $T_{tr} \approx 12$~K at which a bulk magnetic phase transition was inferred from the magnetic susceptibility and specific heat data above. We suspect that the muons are sensing short range ordering at temperatures well above the long range magnetic ordering occurs in PrIrSi$_3$. Further, the $T$ dependences of the fitting parameters $A_{1}$, $A_{2}$, $H_{\rm int}$, $\lambda_1$ and $\lambda_2$ clearly show that all these are nearly temperature independent over $T \leq T_{tr}$ distinguishing the possible long range ordered phase below $T_{tr}$ and short range ordered phase in $T_{tr} < T \leq 30$~K\@.

\section{\label{Sec:INS} Inelastic neutron Scattering}

\begin{figure}
\begin{center}
\includegraphics[width=3.3in]{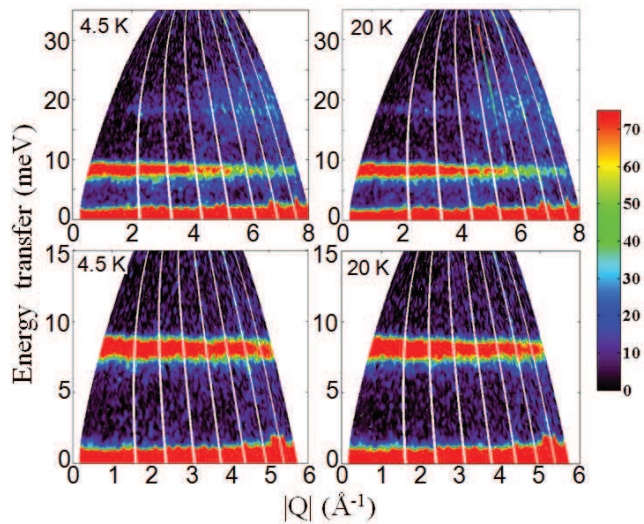}
\caption{\label{fig:INS-PrIrSi3} (Color online) The color-coded contour map of inelastic neutron scattering intensity of PrIrSi$_{3}$ measured with incident energy $E_{i}$ = 20~meV (bottom) and 40~meV (top) at 4.5~K and 20~K plotted as a function of energy transfer $E$ and wave vector transfer $|Q|$. The intensity is in units of mb/sr\,meV\,f.u.}
\end{center}
\end{figure}

\begin{figure}
\begin{center}
\includegraphics[width=2in]{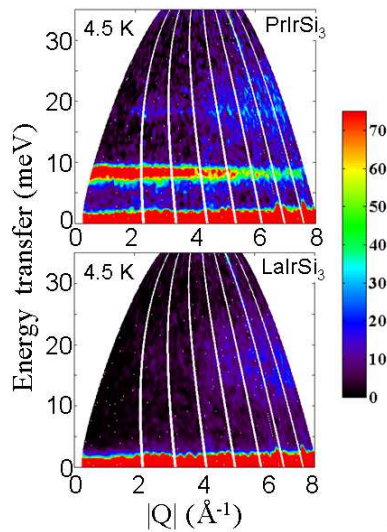}
\caption{\label{fig:INS-PrLaIrSi3} (Color online) The color-coded contour map of inelastic neutron scattering intensity of PrIrSi$_{3}$ (top) and LaIrSi$_{3}$ (bottom) measured with incident energy $E_{i}$ = 20~meV at 4.5~K plotted as a function of energy transfer $E$ and wave vector transfer $|Q|$. The intensity is in units of mb/sr\,meV\,f.u.}
\end{center}
\end{figure}

The INS spectra of PrIrSi$_{3}$ collected with neutrons of incident energies $E_{i} = 20$ and 40~meV for scattering angles between 3$^{\circ}$ and 135$^{\circ}$ at 4.5~K and 20~K are shown in figure~\ref{fig:INS-PrIrSi3} as a color-coded map of the intensity, energy transfer versus momentum transfer. The INS data have been corrected for the background signal and the data have been put into the absolute unit (mb/sr/meV) by using vanadium standard measured with the identical conditions. The color-coded intensity map spectra of PrIrSi$_{3}$ and LaIrSi$_{3}$ obtained with $E_{i} = 40$~meV at 4.5~K are presented in figure~\ref{fig:INS-PrLaIrSi3}.

Two sharp inelastic excitations near 8.5~meV and 18.5~meV are clearly observed in the INS spectra of PrIrSi$_{3}$ at 4.5~K which remain nearly invariant from 4.5~K to 50~K\@. The temperature invariance of inelastic excitations suggests that they result from the crystal field effects. Thus, INS spectra do not show any additional excitation arising from the spin wave contribution due to the magnetic ordering below the transition temperature. Further, no magnetic Bragg peak is observed in $Q$-cuts (integrated over the elastic line from $-2$~meV to 2~meV) at 4.5~K\@. This may suggest that the ordered state moment of the Pr ion is too weak to be detected in the present data.

\begin{figure}
\begin{center}
\includegraphics[width=3in, keepaspectratio]{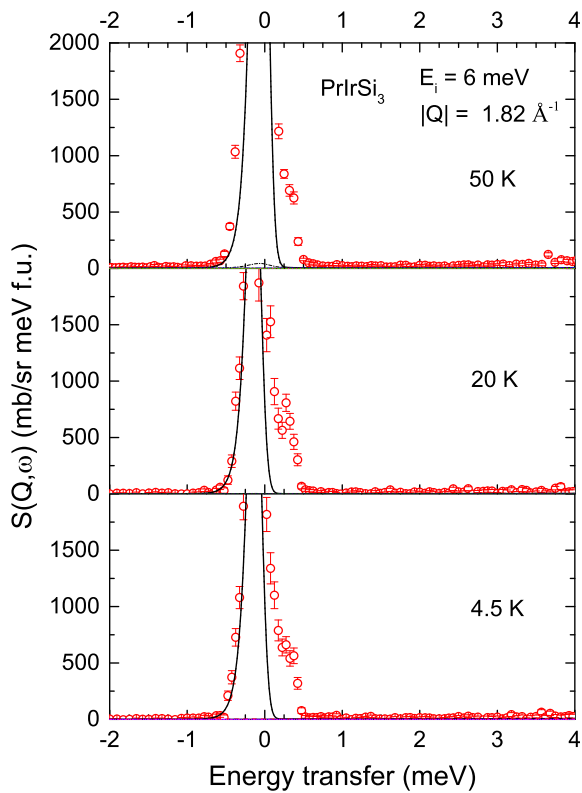}
\caption{\label{fig:INS-Smag1} (Color online) The estimated magnetic scattering from PrIrSi$_{3}$ at different temperatures for momentum transfer $|Q| = 1.82$~{\AA}$^{-1}$ measured with incident energy $E_{i}= 6$~meV. The solid lines are the fits based on the crystal electric field model and the dashed and dash-dotted lines are the components of the fit. Note: the temperature independent scattering on neutron energy loss side is the background from the closed cycle refrigerator.}
\end{center}
\end{figure}

\begin{figure}
\begin{center}
\includegraphics[width=3in, keepaspectratio]{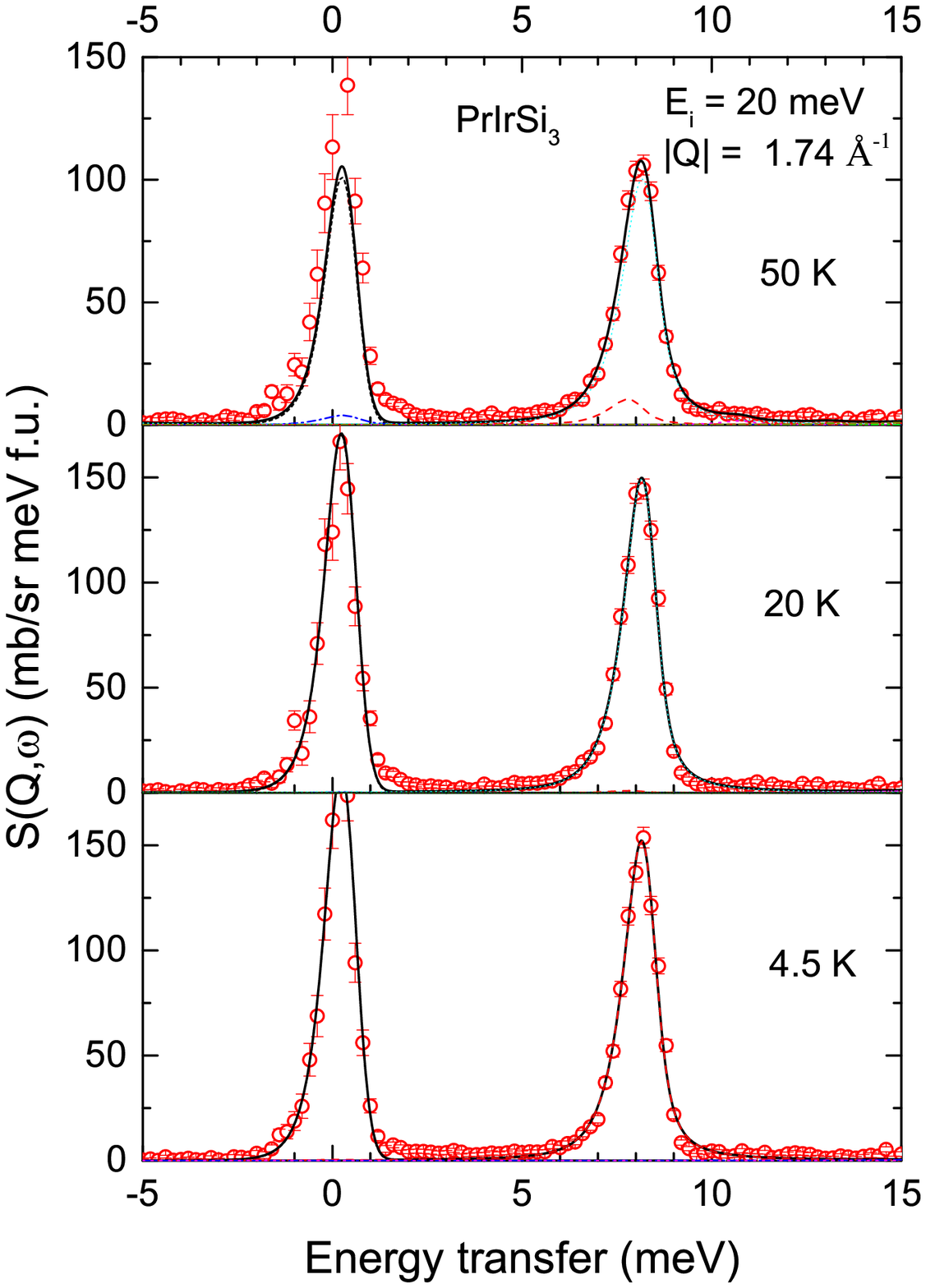}
\caption{\label{fig:INS-Smag2} (Color online) The estimated magnetic scattering from PrIrSi$_{3}$ at different temperatures for momentum transfer $|Q|= 1.74$~{\AA}$^{-1}$ measured with incident energy $E_{i}= 20$~meV. The solid lines are the fits based on the crystal electric field model and the dashed and dash-dotted lines are the components of the fit.}
\end{center}
\end{figure}

\begin{figure}
\begin{center}
\includegraphics[width=3in, keepaspectratio]{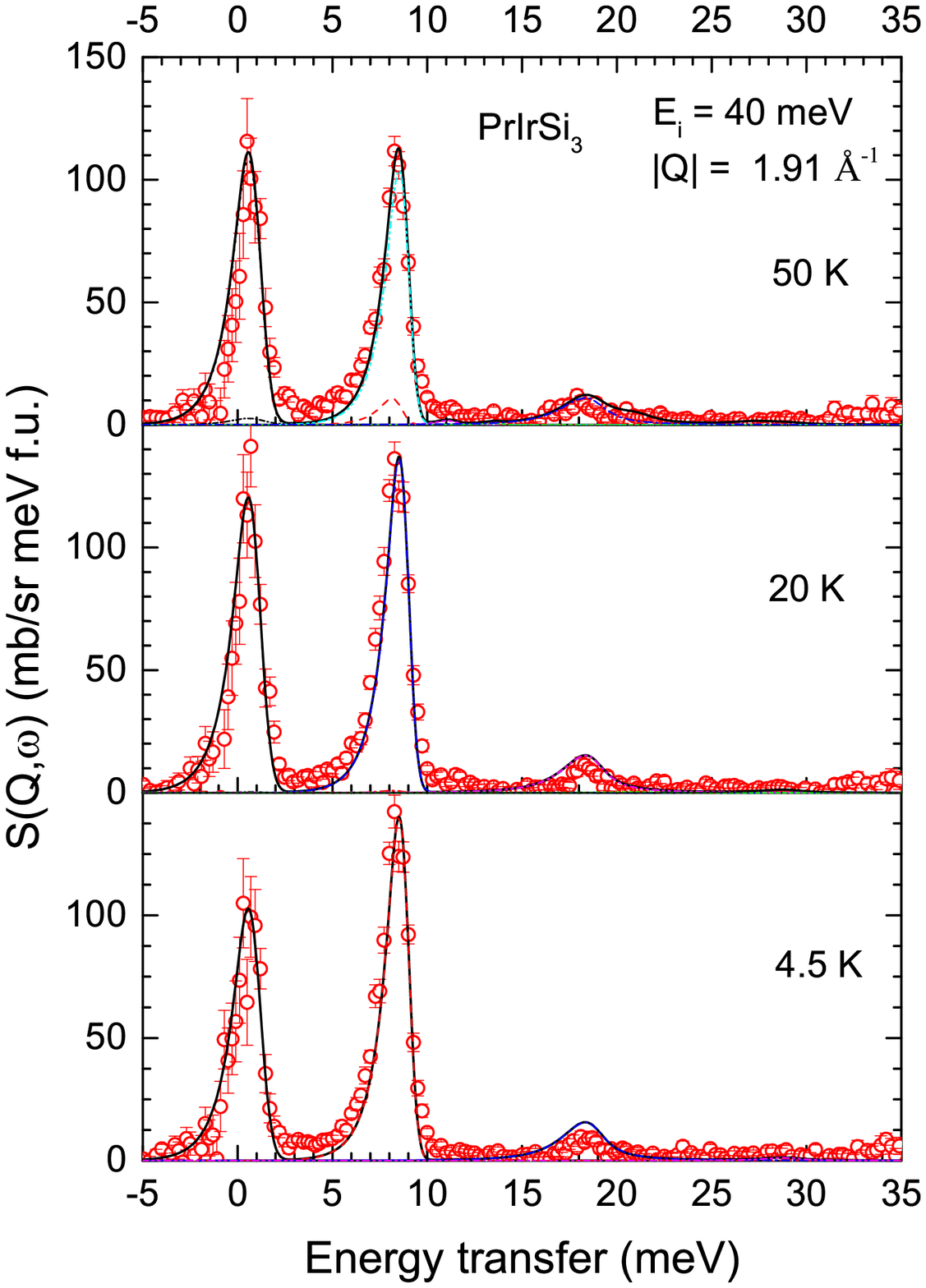}
\caption{\label{fig:INS-Smag3} (Color online) The estimated magnetic scattering from PrIrSi$_{3}$ at different temperatures for momentum transfer $|Q|=1.91$~{\AA}$^{-1}$ measured with incident energy $E_{i}= 40$~meV. The solid lines are the fits based on the crystal electric field model and the dashed and dash-dotted lines are the components of the fit.}
\end{center}
\end{figure}

\begin{figure}
\begin{center}
\includegraphics[width=8cm, keepaspectratio]{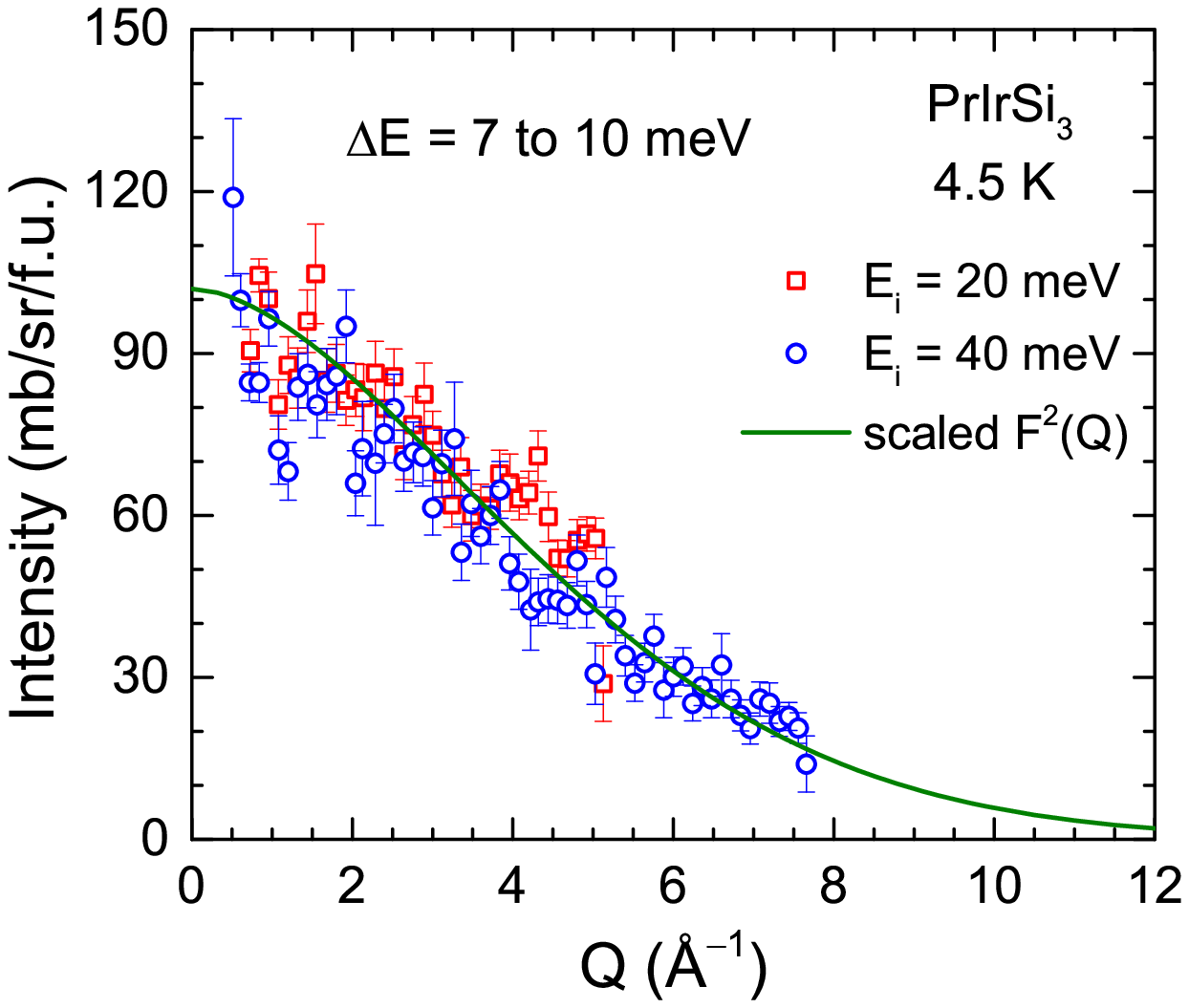}
\caption{\label{fig:INS-form-factor} (Color online) The $Q$ dependence of total intensity of PrIrSi$_{3}$ integrated between 7~meV and 10~meV at 4.5~K for incident energy $E_{i}$ = 20~meV and 40~meV. The solid line represents the square of the Pr$^{3+}$ magnetic form factor, scaled to 102 at $Q = 0$.}
\end{center}
\end{figure}

The dynamic structure factor $S(Q,\omega)$ of neutron scattering of unpolarized neutrons is given by, \cite{Moritz}
\begin{equation}
S(Q,\omega) = \frac{\hbar \omega }{1-\exp(-\hbar \omega /k_{\rm B} T)}F^2(Q)\chi_0 P(Q, \hbar \omega)
\label{eq:S}
\end{equation}
where $F(Q)$ is the magnetic form factor of Pr$^{3+}$, $\chi_0$ is the static bulk susceptibility and $P(Q, \hbar \omega)$ is the normalized spectral function (usually Lorentzian). The static susceptibility is given by
\begin{equation}
\chi_0  = \sum_m \chi_{\rm C}^m + \sum_{m\ne n} \chi_{\rm VV}^{mn}
\label{eq:Chi-0}
\end{equation}
with
\begin{equation}
\chi_{\rm C}^m = \frac{g_J^2 \mu_{\rm B}^2}{k_{\rm B} T} \rho_m \sum_{\alpha = x,y,z} |\langle m |J^\alpha|m \rangle|^2
\label{eq:Chi-C}
\end{equation}
and
\begin{equation}
\chi_{\rm VV}^{mn} = 2 g_J^2 \mu_{\rm B}^2 (\rho_n - \rho_m) \frac{\sum_{\alpha = x,y,z} |\langle m |J^\alpha|n \rangle|^2}{\Delta_{mn}}.
\label{eq:Chi-VV}
\end{equation}
Here $\rho_m = \exp(-E_m/k_{\rm B} T)/Z$ is the population of CEF level with wave function $|m \rangle$ and energy eigenvalue $E_m$, $Z$ is the partition function, $J^\alpha$ is the $x$, $y$ or $z$ component of angular momentum operator, and $\Delta_{mn} = E_n - E_m$ is the energy difference between the CEF levels $|m \rangle$ and $|n \rangle$.

The magnetic scattering of PrIrSi$_{3}$ for different incident energies estimated by subtracting off the phonon contributions using the scattering from LaIrSi$_{3}$, $S(Q,\omega)_{\rm mag} =    S(Q,\omega)_{\rm PrIrSi_3} - \alpha\,S(Q,\omega)_{\rm LaIrSi_3}$ is shown in figures~\ref{fig:INS-Smag1},~\ref{fig:INS-Smag2} and \ref{fig:INS-Smag3}. The scaling factor $\alpha =0.778$ is the ratio of the scattering cross sections of PrIrSi$_{3}$ and LaIrSi$_{3}$. The $S(Q,\omega)_{\rm mag}$ shown in figure~\ref{fig:INS-Smag3} clearly demonstrates the observation of two inelastic excitations centered around 8.5~meV and 18.5~meV. The $Q$-dependent integrated intensity between 7~meV and 10~meV at 4.5~K is shown in figure~\ref{fig:INS-form-factor}. The integrated intensity follows the square of Pr$^{3+}$ magnetic form factor $F^{2}(Q)$ suggesting that the inelastic excitation results mainly from single-ion CEF-transitions. It is to be noted that due to very weak intensity of the 18.5~meV peak and also due to the presence of phonon peak at the same energy as seen in high-$Q$, it was not possibile to analyze the $Q$-dependent intensity of this peak. Further support of magnetic origin of the 18.5~meV peak comes from an initial decrease (at low-$Q$) of the energy integrated (between 17~meV and 21~meV) $Q$-dependence of the intensity.

\begin{table}
\caption{\label{tab:CEF} Crystal field parameters $B_{n}^{m}$ and splitting energies $\Delta_i$ of excited states (with respect to ground state, $\Delta_0 \equiv 0$) obtained from the refinement of the INS data of PrIrSi$_{3}$.}
\begin{ruledtabular}
\begin{tabular}{cc}
$B_2^0$ (meV) & $+0.30(5)$  \\
$B_4^0$ (meV) & $+0.77(2)\times 10^{-2}$  \\
$B_4^4$ (meV) & $-0.677(9)\times 10^{-1}$ \\
$B_6^0$ (meV) & $-0.160(6) \times 10^{-3}$ \\
$B_6^4$ (meV) & $-0.127(6)\times 10^{-2}$ \\
$\Delta_1$ (meV) & 7.95 \\
$\Delta_2$ (meV) & 15.53 \\
$\Delta_3$ (meV) & 17.96 \\
$\Delta_4$ (meV) & 28.48 \\
$\Delta_5$ (meV) & 34.78 \\
$\Delta_6$ (meV) & 46.53 \\
\end{tabular}
\end{ruledtabular}
\end{table}

\begin{figure}
\begin{center}
\includegraphics[width=3in, keepaspectratio]{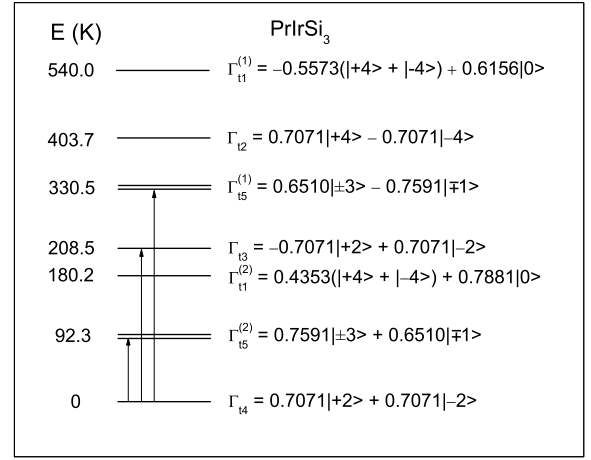}
\caption{\label{fig:CEF-level} (Color online) Crystal electric field level scheme and $f$-electron wave functions of Pr$^{3+}$ ions in PrIrSi$_{3}$ deduced from the analysis of inelastic neutron scattering data. The transitions from the ground state to the excited states that contribute to the observed excitations are shown by arrows.}
\end{center}
\end{figure}

The INS data were analyzed by a model based on crystal electric field effect which fits the INS data very well confirming that the two sharp inelastic excitations observed near 8.5~meV and 18.5~meV arise from the crystal field effect. The crystal field Hamiltonian for a rare earth ion in a tetragonal symmetry (point group $C_{4v}$) is given by
\begin{equation}\label{H-CEF}
 H_{CEF} = B_{2}^{0}O_{2}^{0} + B_{4}^{0}O_{4}^{0} + B_{4}^{4}O_{4}^{4} + B_{6}^{0}O_{6}^{0} + B_{6}^{4}O_{6}^{4}
\end{equation}
where $B_{n}^{m}$ are the phenomenological crystal field parameters and $O_{n}^{m}$ are the Stevens operators. The crystal electric field removes the nine-fold degeneracy of $4f$ ground multiplet of Pr$^{3+}$ ($J =4$) leading to five singlets ($\Gamma_{t1}^{(1)}$, $\Gamma_{t1}^{(2)}$, $\Gamma_{t2}$, $\Gamma_{t3}$, $\Gamma_{t4}$) and two doublets ($\Gamma_{t5}^{(2)}$, $\Gamma_{t5}^{(1)}$) in tetragonal symmetry, where the $\Gamma_{j}$'s are the irreducible representations of the point group. The wave functions for  $\Gamma_{j}$ are shown in figure~\ref{fig:CEF-level}.

In CEF analysis it is very important to find a good starting set of the CEF parameters.  We have used Monte-Carlo method given in FOCUS program \cite{Fabi1995}, which allowed us to find the best starting value of the CEF parameters. Then we simultaneously fitted all different $E_{i}$ and temperatures data sets. Our simultaneous fit of nine sets of INS data of PrIrSi$_{3}$ for $E_i = 6$~meV, 20~meV and 40~meV at 4.5~K, 20~K and 50~K are shown by solid lines in figures~\ref{fig:INS-Smag1},~\ref{fig:INS-Smag2} and \ref{fig:INS-Smag3}. A nice agreement between the INS data and the fit is seen from figures~\ref{fig:INS-Smag1},~\ref{fig:INS-Smag2} and \ref{fig:INS-Smag3}. The phenomenological crystal field parameters $B_{n}^{m}$ obtained from the simultaneous fit are listed in Table~\ref{tab:CEF}. The CEF wave functions and the CEF level scheme corresponding to the $B_{n}^{m}$ parameters in Table~\ref{tab:CEF} are shown in figure~\ref{fig:CEF-level}. The transitions from the ground state to the excited states that contribute (has nonzero matrix element) to the observed excitations are shown by arrows in figure~\ref{fig:CEF-level}. The ground state is a singlet lying below an excited doublet at 92.3~K\@.

\begin{figure}
\begin{center}
\includegraphics[width=3in, keepaspectratio]{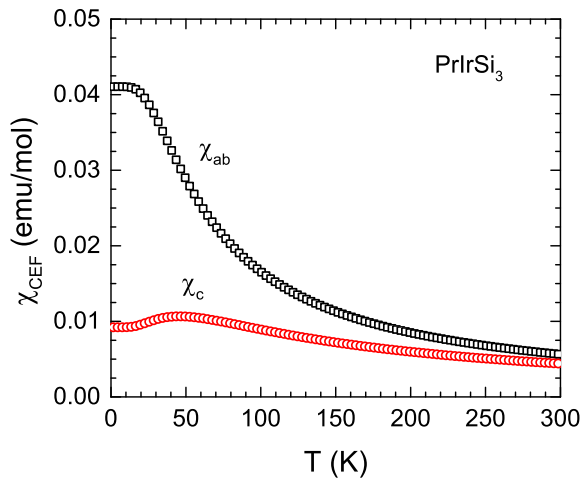}
\caption{\label{fig:INS-Chi} (Color online) The calculated single crystal susceptibility $\chi(T)$ of PrIrSi$_{3}$ using CEF parameters given in Table~\ref{tab:CEF}.}
\end{center}
\end{figure}

The Schottky contribution to specific heat due to CEF can be estimated from \cite{Gopal1966}
\begin{eqnarray}
\label{C_SCh}
C_{\rm Sch}(T) & = & \left(\frac{R}{T^2}\right) \bigg\{ \sum_{i}g_i e^{-\Delta_i/T} \sum_{i}g_i \Delta_i^2 e^{-\Delta_i/T}\\
            & &  - \bigg[\sum_{i}g_i \Delta_i e^{-\Delta_i/T}\bigg]^2  \bigg\} \times \bigg[\sum_{i}g_i e^{-\Delta_i/T}\bigg]^{-2}\nonumber
\end{eqnarray}
where $g_i$ are the degeneracies of CEF levels with energies $\Delta_i$. The CEF specific heat corresponding to the CEF parameters in Table~\ref{tab:CEF} calculated according to Eq.~(\ref{C_SCh}) using the software AMOS \cite{AMOS} is shown in figure~\ref{fig:HC}(b) which shows a nice agreement with the $C_{\rm mag}(T)$ data for $T > T_{tr}$, thus supporting the CEF level scheme deduced from the INS data. The single crystal $\chi_{\rm CEF}(T)$ calculated according to Eq.~(\ref{eq:Chi-0}) using the software AMOS is shown in figure~\ref{fig:INS-Chi}. The magnitude of $\chi_{\rm CEF}(T)$ is much smaller than the measured $\chi(T)$ in figure~\ref{fig:Chi} (can be attributed to the presence of ferromagnetic correlation), however $\chi_{\rm CEF}(T)$ reveals strong anisotropy, $\chi_{ab} > \chi_{c}$ over the entire temperature range up to 300~K, revealing an easy plane behavior of magnetic moments.

\section{\label{Discussion} Discussion}

The electronic and magnetic properties of PrIrSi$_{3}$ indicate that it shows a magnetic phase transition with a singlet ground state. The inelastic neutron scattering data, $C_{\rm mag}(T)$ and $S_{\rm mag}(T)$ consistently reveal a singlet ground state in PrIrSi$_{3}$. On the other hand $\chi(T)$ and $C_{\rm p}(T)$ reveal a magnetic phase transition at $T_{tr} \approx 12$~K. Magnetic phase transition is further inferred from $\mu$SR results which show magnetic ordering at an elevated temperature of 30~K, possibly due to the presence of short range correlations well above the long range magnetic ordering inferred from the specific heat and magnetic measurements. The nature of this magnetic ordering in PrIrSi$_{3}$ is not clear. However, considering the positive $\theta_{\rm p}$, irreversibility between the ZFC and FC $\chi$, and very slow relaxation of $M(t)$, dominance of ferromagnetic correlations is evident.

Many singlet ground state systems are known to exhibit long range magnetic ordering due to the exchange-generated admixture of ground state singlet and the excited CEF levels to meet the criterion of induced moment magnetism based on the relative strength of exchange interaction and splitting energy \cite{Trammel1963,Cooper1967,Wang1968, Bleaney1963}. However, the crystal field splitting of 92~K between the ground state and the first excited state doublet of PrIrSi$_{3}$ is very high therefore a very large exchange energy is required for an induced moment magnetic ordering in this compound. The singlet ground state system PrRu$_2$Si$_2$ crystallizing with body-centered tetragonal structure orders ferromagnetically below $T_{\rm c} = 14$~K with splitting energy of 26~K (2.25~meV) between the ground state singlet and first excited state singlet \cite{Mulders1997}. PrNi$_2$Si$_2$ is another singlet ground state system with body-centered tetragonal structure that exhibits amplitude modulated long range antiferromagnetic order below $T_{\rm N} = 20$~K and has first excited singlet at 38~K (3.3~meV) \cite{Blanco1992,Blanco1997,Blanco2013}. Both PrRu$_2$Si$_2$ and PrNi$_2$Si$_2$ exhibiting long range induced moment magnetic ordering with singlet ground state system have low CEF splitting energies, therefore exchange-generated admixture of low-lying crystal field levels is achieved. In contrast, the admixture of low-lying crystal field levels by exchange interaction in PrIrSi$_{3}$ with large CEF splitting energy of 92~K is unexpected and the observation of magnetic ordering in this singlet ground state system is not easy to understand.

Induced moment magnetism has also been reported in singlet ground state systems with splitting energy larger than that of PrRu$_2$Si$_2$ and PrNi$_2$Si$_2$ but somewhat smaller than that of  PrIrSi$_{3}$.  Cubic structured singlet ground state compounds Pr$_3$Tl and Pr$_3$In exhibit long range induced moment magnetic ordering with  $T_{\rm c} = 11.6$~K (ferromagnetic) and $T_{\rm N} = 12$~K (antiferromagnetic), respectively \cite{Birgeneau1971,Andres1972,Christianson2005,Christianson2007}. They both have triplet as excited state with splitting energies of 79~K (6.8~meV) and 73~K (6.3~meV), respectively \cite{Birgeneau1971,Christianson2007}. Face centered cubic (fcc) Pr also shows an induced moment ferromagnetism ($T_{\rm c} = 20$~K) with larger splitting energy of 84~K (7.2~meV) between singlet ground state and triplet excited state \cite{Birgeneau1971,Cooper1972,Bucher1969}. Orthorhombic PrCoAl$_4$ with singlet ground state is reported to exhibit an amplitude modulated antiferromagnetic ordering below $T_{\rm N} = 17$~K possibly with a large splitting energy of about 100~K between the ground state and the first excited state as deduced from the specific heat data, that remains to be verified by inelastic neutron scattering \cite{Tung2004,Papamantellos2004}.

For a singlet ground system like PrIrSi$_{3}$ the mean-field critical temperature for self-induced moment ordering is given by \cite{Goremychkin2008}
\begin{equation}
T_{c} =  \Delta \left\{ \ln \left[ \frac{\mathcal{J}_{\rm ex} \alpha^{2} + n \Delta}{\mathcal{J}_{\rm ex} \alpha^{2} - \Delta}\right] \right\}^{-1}
\label{eq:Tc}
\end{equation}
\noindent where $\alpha$ is the matrix element between the ground state singlet and the excited state doublet and $n$ is the degeneracy of the excited state which for a doublet is $n =2$. Our analysis of INS data gives $\alpha^2$ = 10.46 and $\Delta$ = 7.95~meV, therefore according to Eq.~\ref{eq:Tc}, in order to have an induced moment ordering at 12.2 K the critical exchange required is $\mathcal{J}_{\rm ex} = 0.761$~meV.

On the other hand the exchange energy estimated from $\theta_{\rm p}$ according to
\begin{equation}
\theta_{\rm p} = -\frac {\mathcal{J}_{ex} J(J+1)}{3 k_{\rm B} }
\label{eq:Jex}
\end{equation}
with $\mathcal{J}_{\rm ex} =  \sum_{j}\mathcal{J}_{ij} $, and $k_{\rm B}$ is Boltzmann's constant, is $\mathcal{J}_{\rm ex} = -0.336$~meV for $\theta_{\rm p} = +26(2)$~K as determined in Sec.~\ref{Sec:Magnetization} from $\chi(T)$. This value of $|\mathcal{J}_{\rm ex}| = 0.336$~meV is much smaller than the $\mathcal{J}_{\rm ex} = 0.761$~meV required. Even though a more reliable estimate of $\mathcal{J}_{\rm ex}$ is necessary, it is apparent from the $\mathcal{J}_{\rm ex}$ values that there should be no spontaneous induced moment ordering in PrIrSi$_{3}$. Thus the observation of magnetic ordering in this singlet ground state PrIrSi$_{3}$ is not understood on the basis of relative strength of $\mathcal{J}_{\rm ex}$ and $\Delta$.

\section{\label{Conclusion} Conclusions}

The physical properties of PrIrSi$_{3}$ which forms in BaNiSn$_{3}$-type tetragonal crystal structure (space group $I4\,mm$) are reported.  A magnetic phase transition at $T_{tr} \approx 12$~K of ferromagnetic nature considering the irreversibility between the ZFC and FC susceptibility and very slow relaxation of thermoremnant magnetization is inferred from the magnetic susceptibility and specific heat data. Similar irreversibility in $\chi(T)$, strong dependence of $\chi$ on applied $H$ and slow relaxation of $M(t)$ have also been observed in spin-glass systems PrRuSi$_3$ and PrRhSn$_3$ \cite{Anand2011,Anand2012a}. However, the ac magnetic susceptibility and $\mu$SR measurements do not show signatures of spin-glass behavior in PrIrSi$_{3}$.

Presence of strong crystal field effect is revealed by a broad Schottky-type anomaly in the magnetic part of specific heat and a broad curvature in electrical resistivity. A CEF-split singlet ground state is inferred from the extremely small value of magnetic entropy below $T_{tr}$. The INS data confirms the CEF-split singlet ground state in PrIrSi$_{3}$. Two sharp inelastic excitations are observed at 8.5~meV and 18.5~meV in INS spectra. The INS data are well explained by a model based on crystal electric field effect. The crystal field level scheme and wave functions have been evaluated. A large CEF splitting energy of 92~K between the ground state singlet and the first excited doublet is found from the INS data consistent with the magnetic specific heat.

The $\mu$SR data reveal possible magnetic ordering below 30~K which is much higher than the $T_{tr}$ possibly due to the presence of short range correlations well above the long range magnetic ordering. Thus $\chi(T)$, $C_{\rm p}(T)$ and $\mu$SR provide conclusive evidence for magnetic ordering in PrIrSi$_{3}$. However, considering the large splitting energy of 92~K the magnetic ordering in singlet ground state PrIrSi$_{3}$ is unexpected. Very recently we also observed a similar magnetic behavior below 15~K in singlet ground state PrRhSi$_3$ which also has a large CEF splitting energy of 81~K between the ground state singlet and the first excited doublet as inferred from the heat capacity analysis \cite{Anand2013}. The difference between these two systems is that the specific heat of PrRhSi$_3$ does not show a corresponding anomaly like PrIrSi$_{3}$. The observation of long range magnetic ordering in singlet ground state PrIrSi$_{3}$ is enigmatic and needs further investigations. It should be noted that the heat capacity, resistivity  and  magnetic susceptibility do not show any sign of ordering at 30~K, where $\mu$SR reveals collective behavior that is of putative magnetic origin. This may also suggest that the implanted muons in PrIrSi$_{3}$ are playing an important role in the 30~K phase transition. This type of muon induced magnetic ordering has been seen in PrNi$_5$, PrIn$_3$ and Pr$_2$Rh$_3$Ge$_5$ \cite{Feyerherm1995,Tashma1997,Anand2014}. Investigations on single crystals of PrIrSi$_{3}$ are highly desirable to understand the observed complex magnetism.

\acknowledgments

Authors VKA, DTA and ADH would like to acknowledge financial assistance from CMPC-STFC grant number CMPC-09108. AB would like to thank UJ and STFC for PDF funding. AMS thanks the URC of UJ, and the SA-NRF (78832) for financial assistance.

\end{document}